\PassOptionsToPackage{pdfpagelabels=false}{hyperref}
\documentclass[twocolumn]{aastex63}

\usepackage{silence}
\usepackage{natbib}
\usepackage{times}
\usepackage{enumitem}
\usepackage{graphicx}
\usepackage{amsmath}
\usepackage{bm}
\usepackage{graphics}
\usepackage{url}
\usepackage{epsfig}
\usepackage{ulem}
\usepackage{verbatim}
\usepackage{float}
\usepackage{lastpage}
\usepackage{amsmath}

\newcommand{\frb}{FRB\,180916}

\def\lsim{\hbox{\rlap{\raise 0.425ex\hbox{$<$}}\lower 0.65ex\hbox{$\sim$}}}
\def\gsim{\hbox{\rlap{\raise 0.425ex\hbox{$>$}}\lower 0.65ex\hbox{$\sim$}}}
\def\arcmin{\hbox{$^\prime$}}
\def\arcsec{\hbox{$^{\prime\prime}$}}

\shorttitle{Optical Observations of FRB\,180916}
\shortauthors{Kilpatrick et al.}

\begin{document}
\sloppy

\title{Deep optical observations contemporaneous with emission from the periodic FRB 180916.J0158+65}

\correspondingauthor{C. D. Kilpatrick}
\email{ckilpatrick@northwestern.edu}

\newcommand{\NU}{\affiliation{Center for Interdisciplinary Exploration and Research in Astrophysics (CIERA) and Department of Physics and Astronomy, Northwestern University, Evanston, IL 60208, USA}}

\newcommand{\NMSU}{\affiliation{Astronomy Department, Box 30001, Department 4500, New Mexico State University, Las Cruces, NM 88003-0001}}

\newcommand{\UCSC}{\affiliation{Department of Astronomy and Astrophysics, University of California, Santa Cruz, CA 95064, USA}}

\newcommand{\APO}{\affiliation{Apache Point Observatory and New Mexico State University, P.O. Box 59, Sunspot, NM, 88349-0059, USA}}

\newcommand{\UCB}{\affiliation{Astronomy Department and Theoretical Astrophysics Center, University of California, Berkeley, Berkeley, CA 94720, USA}}

\newcommand{\UI}{\affiliation{Centre for Astrophysics and Cosmology, Science Institute, University of Iceland, Dunhagi 5, 107 Reykjav\'{i}k, Iceland}}

\newcommand{\Pont}{\affiliation{
Instituto de F\'{i}sica, Pontificia Universidad Cat\'{o}lica de Valpara\'{i}so, Casilla 4059, Valpara\'{i}so, Chile}}
\author[0000-0002-5740-7747]{Charles D. Kilpatrick}
\NU

\author{Joseph N. Burchett}
\NMSU

\author{David O. Jones}
\UCSC

\author{Ben Margalit}
\UCB

\author{Russet McMillan}
\APO

\author{Wen-fai Fong}
\NU

\author{Kasper E. Heintz}
\UI

\author{Nicolas Tejos}
\Pont

\author{Alicia Rouco Escorial}
\NU

\begin{abstract}

We present deep Apache Point Observatory optical observations within seconds of the outburst of the periodic fast radio burst (FRB) 180916.J0158+65 obtained on 3 September 2020. FRB\,180916.J0158+65 is located in a nearby spiral galaxy 150~Mpc away and has an ``active phase'' with a well-measured period of approximately 16.3~days. Targeting the FRB at the peak of its expected active phase and during a recent 30~minute observing window by the Canadian Hydrogen Intensity Mapping Experiment (CHIME) in which a radio burst was detected, we did not detect any transient optical emission at $m_{i}\approx$24.7~mag (3$\sigma$) from 2.2 to 1938.1~seconds after the burst arrival time in optical bands (corrected for dispersion). Comparing our limiting magnitudes to models of a synchrotron maser formed in the circumburst environment of FRB\,180916+J0158.65, we constrain scenarios where the burst energy was $>10^{44}$~erg and the circumburst density was $>$10$^{4}$~cm$^{-3}$.

\end{abstract}



\section{Introduction}

Fast radio bursts (FRBs) are millisecond timescale \citep{Lorimer07} bursts of MHz--GHz radio emission from extragalactic sources \citep[][and references therein]{Thornton13,CHIME19,Cordes19,Petroff19}. The detection of the first repeating FRB\,121102 \citep{Spitler14,Spitler16} enabled its precise localization in a host galaxy at $z=0.193$ \citep{Chatterjee17}.  There may also be a non-repeating population of FRBs \citep[e.g.,][]{Ai20,Hashimoto20}, but the observed FRB rate and luminosity function suggests that most of these sources must be repeating \citep[whether or not they are observed to repeat, e.g.,][]{Ravi19,Caleb19}. A handful of these FRBs with multiple detections have been accurately localized and securely associated with host galaxies, spanning a wide range of host types from star forming \citep[FRB~121102 and FRB~190711;][]{Chatterjee17,Macquart20,Li20,Kumar20} to quiescent \citep[FRB~180916 and FRB~180924 with a star-formation rate $<2~M_{\odot}$~yr$^{-1}$ at $z=0.3214$;][]{Bannister19,Marcote20} galaxies \citep[see also][]{Bhandari18,Heintz20}.

Several hypotheses have been proposed for FRB progenitor systems \citep{Platts19}, from eruptions on the surfaces of highly-magnetized neutron stars \citep[i.e., magnetars;][]{Popov13,Lyubarsky14,Kulkarni14,Katz16,Beloborodov17,Kumar17,Metzger17,Wadiasingh19,Beniamini20} to accretion-induced collapse of neutron stars into black holes \citep{Falcke14}. Notably, repeating FRBs exhibit pulse widths $>$4$\sigma$ longer than those without multiple detected bursts \citep{CHIME19}, suggesting that there may be multiple progenitor channels.  In addition, the recent detection of an FRB-like event from the Galactic magnetar SGR\,1935+2154 suggests that at least a subset of repeating, extragalactic FRBs originate from magnetars \citep[][]{frb200428,Bochenek20,Lu20,Margalit20b}. This interpretation is complicated in part by evidence that FRB hosts lack a clear association to star formation \citep[although some studies suggest FRBs can be produced via magnetars that do not trace star formation; see, e.g.,][]{Margalit19b,Safarzadeh20,Bochenek20}.

Precise localization of FRBs has also enabled both targeted and untargeted follow up at wavelengths spanning from optical \citep{Hardy17,Bhandari18,Andreoni20} to X-ray \citep{Petroff15,Scholz16,Pilia20,Tavani20,Scholz20} to gamma-ray wavelengths \citep{Yamasaki16,Best19,Cunningham19,Guidorzi19,Guidorzi20}, primarily for the well-localized, repeating bursts FRB\,121102 and FRB\,180916.  So far these searches have revealed no potential counterparts for extragalactic FRBs \citep[e.g.,][]{Chen20} apart from a 100~s duration gamma-ray transient detected at the 3.2$\sigma$ level by the Neil Gehrels {\it Swift} Observatory \citep[\textit{Swift};][]{Gehrels2004} and contemporaneous with a burst from FRB\,131104 \citep[][]{Delaunay16}, although the association may not be secure \citep{Shannon17,Gao17}.

Theoretical models predict optical emission spanning faint luminosities of $<10^{39}$~erg~s$^{-1}$ within seconds of the burst via synchrotron emission in the circumburst medium of the progenitor \citep[][]{Metzger19}.  However, the best limits from optical follow up observations have only ruled out counterparts to bursts down to relatively unconstraining luminosities of $\approx$10$^{45}$~erg~s$^{-1}$ \citep{Hardy17} within milliseconds of a burst or $\approx$5$\times$10$^{42}$~erg~s$^{-1}$  within minutes of a burst \citep{Andreoni20}.  Recent constraints on the periodic activity window of the repeating FRB\,180916.J0158+65 (hereafter FRB\,180916; \citealt{CHIME20}) offer a unique opportunity to target emission from a FRB counterpart at non-radio wavelengths.  In addition to untargeted optical observations from the Zwicky Transient Facility \citep{Andreoni20}, \citet{Pilia20} and \citet{Zampieri20} reported high-speed optical observations with the 1.2m Galileo telescope during a burst that limit the Sloan $i$-band fluence of  FRB\,180916 to $<$0.151~Jy~ms.  Additional multi-band follow up of this source will further benefit from the known 16.3~day period, suggesting that the counterpart is highly active on this timescale \citep{CHIME20}.  The relatively low redshift \citep[$z=0.0337$ with an implied luminosity distance of $D_{L}\approx150$~Mpc;][]{CHIME20} means that any observations will yield significantly deeper constraints than those for FRBs previously targeted at optical wavelengths \citep[e.g., FRB~121102;][]{Hardy17,Bhandari18}.

Here we discuss targeted optical follow up of FRB\,180916 with the Apache Point Observatory (APO) 3.5m telescope.  These observations were obtained contemporaneously with observations from the Canadian Hydrogen Intensity Mapping Experiment (CHIME) on 3 September 2020.  During the observations, CHIME detected a radio burst  at a location and dispersion measure consistent with previous bursts from FRB\,180916 \citep{CHIME20,Marcote20}.  At the time of the burst and for 30~minutes thereafter, we observed the FRB\,180916 host galaxy with the APO telescope, but we did not detect any transient optical emission at the FRB site.  Based on the non-detections, we place constraints on the allowed burst properties and circumburst density in the synchrotron maser model.  We discuss the timing and details of our observations in \autoref{sec:obs}.  In \autoref{sec:analysis}, we analyze these limits in the context of realistic optical counterparts to FRBs and discuss the implications of these limits for future follow up efforts.  We summarize our findings in \autoref{sec:conclusions}.

\begin{figure*}
    \centering
    \includegraphics[width=\textwidth]{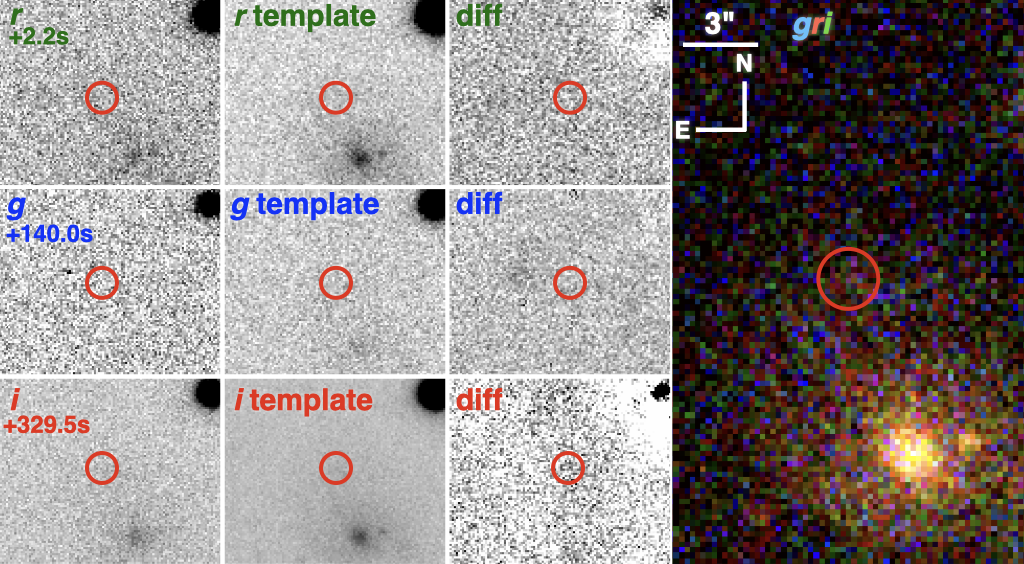}
    \caption{{\it Upper row}: $r$-band image starting from +2.2~s relative to the dispersion-corrected burst arrival time, our $r$-band template from 13 September 2020, and the difference between the two frames.  There is no signature of transient emission at the FRB location (red circles).  {\it Middle row}: Same as the top row for our $g$-band image starting at +140.0~s.  {\it Bottom row}: Same as the top row for our $i$-band image starting at +329.5~s.  {\it Right}: RGB ($igr$) image showing all of our stacked, post-burst data and centered on the location of the burst.}
    \label{fig:frb-image}
\end{figure*}

\section{Observations}\label{sec:obs}

We targeted \frb\ on 3 September 2020 with the APO 3.5m telescope, mounted with the Astrophysical Research Consortium Telescope Imaging Camera \citep[ARCTIC;][]{ARCTIC}. Our observations began (i.e., the camera shutter opened) at UTC 2020-09-03 11:05:39.7503.  Each exposure sequence was a $3\times30.33$~s set of images in a single band (except for the third and fourth set of $g$-band exposures, which consisted of 2 and 4 exposures, respectively), following a $i \rightarrow r \rightarrow g$ pattern for 9 images per pattern or 36 images over the full set of observations (\autoref{tab:observations}).  The average time per exposure is 30.3~s over 2221.7~s for an observing efficiency of $\approx$49\%. In addition, we obtained follow up observations on 13 September 2020 in $gri$ bands with $3\times100$~s exposures to use as templates for comparison to the previous epoch. We show example $r$ and $i$-band images obtained within $2$~min from the topocentric radio burst arrival time in \autoref{fig:frb-image}.

The radio burst occurred during our first $gri$ sequence when the second $r$-band image was exposing, implying that 32 out of 36 of our exposures occurred around or after the burst arrival time. Based on the topocentric burst time at 400~MHz provided by CHIME\footnote{2020-09-03T11:10:32.495964Z; \url{https://www.chime-frb.ca/repeaters/180916.J0158+65}}, the radio burst arrival time was 6.908~s after the shutter opened and 23.442~s before the shutter closed for this exposure. Thus including the second $r$-band exposure, approximately $32\times30.33$~s of cumulative exposure time was obtained during or immediately after the time of burst.  The exact time the shutter opened and closed is given in \autoref{tab:observations} along with the relative time (in seconds) from the radio burst arrival.

The time of arrival for an FRB is frequency ($\nu$) dependent following a $\nu^{-2}$ dispersion law \citep{Tanenbaum68}. Variation in the burst arrival time is used to quantify the dispersion measure (DM) along the line of sight to the burst, a value that correlates with the column of electrons to the source and thus can be used to measure electron densities in the intergalactic medium \citep{Petroff16,Shannon18,Macquart18,Petroff18,Macquart20}. The DM for the FRB\,180916 burst on 3 September 2020 was $352.6\pm3.2$~pc~cm$^{-3}$, consistent with previous measurements toward the repeater \citep{CHIME20,Marcote20}.  Based on this measurement, we can estimate the arrival time for the corresponding $r$-band optical emission from the FRB observed at $400$~MHz, which would be earlier as it occurred at a higher frequency. Following equation (1) in \citet{Cordes19} \citep[see also][]{Tanenbaum68}, we estimate that the burst arrival time in the optical was 9.1~s earlier than at 400~MHz, implying an arrival time of UTC 2020-09-03 11:10:23.4, which we use throughout this paper as the reference point for the ``dispersion-corrected burst arrival time.''

We reduced all ARCTIC data using a custom-built pipeline based on the {\tt photpipe} imaging and photometry package \citep{Rest05,kilpatrick+18}. Each frame was corrected for bias and flat-fielded using bias and sky flat field frames obtained in the same instrumental configuration. We registered the images using 2MASS astrometric standards \citep{Skrutskie06} observed in the field of each image. Finally, we performed point spread function (PSF) photometry using {\tt DoPhot} \citep{Schechter93} and calibrated the $gri$ data using PS1 DR2 standard stars in each image \citep{Flewelling16}.  We subtracted the observations taken on 13 September 2020 from all 3 September 2020 observations using {\tt HOTPANTS} \citep{hotpants} to perform PSF convolution and difference imaging and then estimated the 3$\sigma$ limiting magnitude at the position of FRB\,180916 \citep[from][]{Marcote20} with fake star injection.  Thus the limits we derive can be interpreted as the maximum average in-band specific flux integrated over the 30.33~s window for each exposure or the entire observation window for the limits in the stacked data.  As shown in \autoref{tab:observations}, the typical limiting magnitude of each individual frame was $\approx$24.5~mag in $gri$ bands or $\approx$26.0~mag in the stacked frames.

\section{Constraining the FRB Emission Mechanism}\label{sec:analysis}

\begin{figure}
    \centering
    \includegraphics[width=0.49\textwidth]{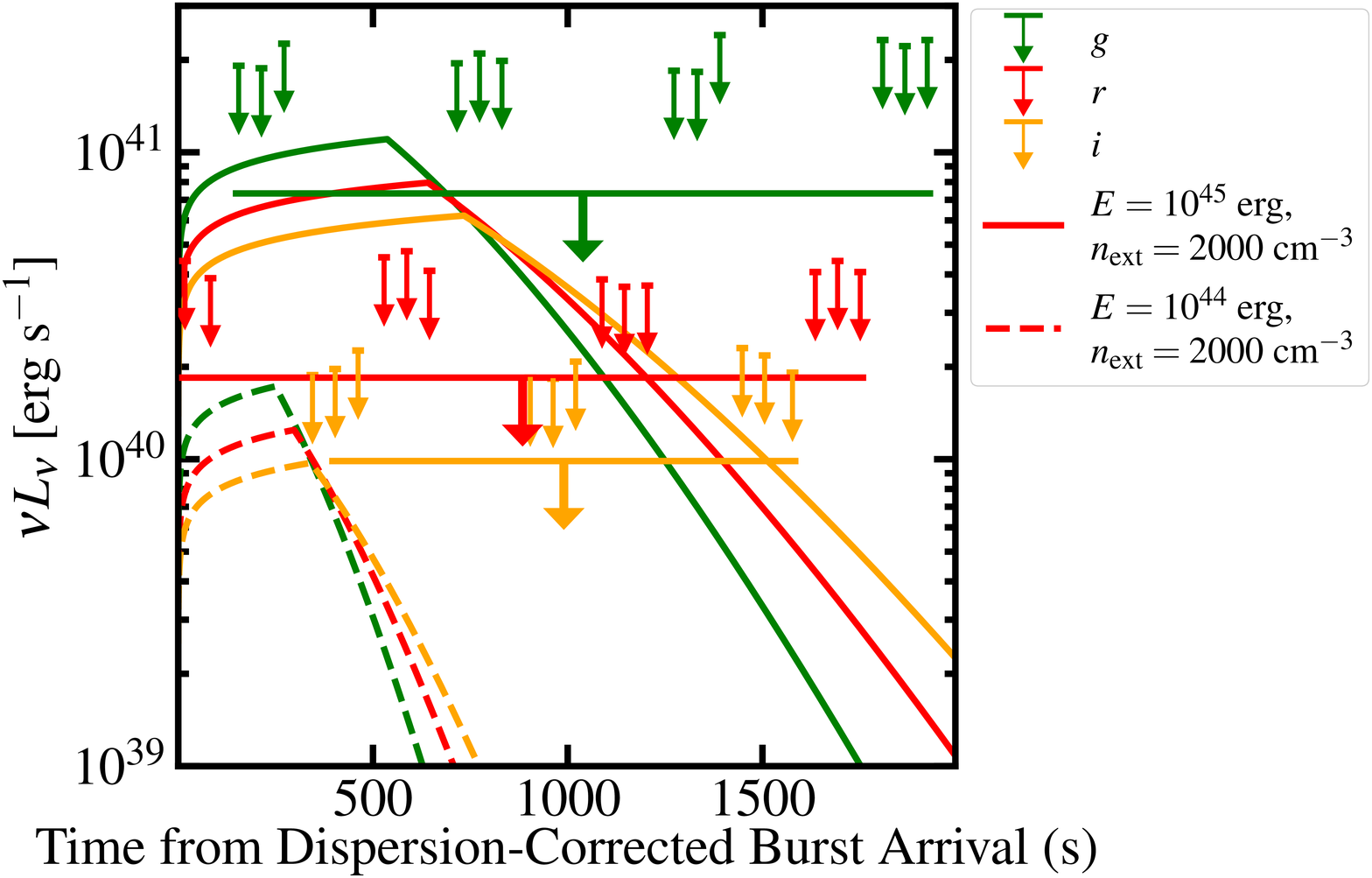}
    \caption{Optical light curves of the synchrotron afterglow from a magnetar-driven burst as described in \autoref{sec:analysis}.  We compare the predicted specific luminosity in $\nu L_{\nu}$ for $r$ band and two sets of model light curves corresponding to injected energies $E=10^{45}$~erg (solid lines) and $E=10^{44}$~erg (dashed lines) in $g$ (green), $r$ (red), and $i$ (orange) to the limits derived from our optical data.  Otherwise, the models assume our fiducial parameters as given in \autoref{sec:analysis} \citep[see also figure 8 in][]{Metzger19} and the median circumburst density ($n_{\mathrm{ext}}\approx2000$~cm$^{-3}$) in \citet{Margalit20a}.  The model is predicted to decline rapidly at the time $t_{\mathrm{syn}}$ where the peak of the synchrotron spectrum cascades below the observing frequency.}
    \label{fig:models}
\end{figure}

The timeline for our exposures immediately after the radio burst is shown in \autoref{fig:models}.  We estimate based on the 9.1~s dispersive delay of the radio emission that the FRB occurred approximately 2.2~s before the shutter opened for our second $r$-band exposure.  Thus our closest and most constraining image of FRB\,180916 covered roughly 2.2 to 30.5~seconds relative to the burst arrival time.  Our full, post-burst data set covers approximately +2.2~s to +1938.1~s relative to the dispersion-corrected burst arrival time in $gri$ (\autoref{tab:observations}).  We also give the limiting magnitudes for times before the expected burst arrival time, but in the context of the models below, we do not consider these data as no optical emission is predicted.

We give the 3$\sigma$ limiting magnitude in AB magnitudes both for a source at the location of FRB\,180916 in each image and in absolute magnitudes after correcting for the distance modulus and Milky Way foreground extinction $A_{V}=2.767$~mag \citep[from][]{Schlafly11}.  We assume the redshift $z=0.0337\pm0.0002$ derived for the host galaxy of FRB\,180916 in \citet{Marcote20} along with \citet{Planck15} cosmology, from which we derive a luminosity distance of $D_{L}=153\pm1$~Mpc or a distance modulus of $\mu=35.92\pm0.02$~mag.

We compare these limits to predictions of the synchrotron maser model as shown in \citet{Metzger19,Margalit20a} \citep[but also see][for slightly different optical predictions, especially at times comparable to the burst duration when the total luminosity may be larger]{Lyubarsky14,Beloborodov17,Beloborodov20}.  In the \citet{Metzger19} formalism for this model, the radio burst originates in a shock from a radially-expanding plasmoid launched from a central engine (e.g., a magnetar). The relativistic plasmoid may be decelerated by surrounding material in the immediate environment of the engine.  If this material is sufficiently magnetized, synchrotron maser emission will be produced \citep{Plotnikov19}.
In the model of \cite{Metzger19,Margalit20a} (and first proposed by \citealt{Beloborodov17}) the surrounding upstream material is baryon-loaded ejecta expelled in previous flaring activity of the magnetar.

One prediction of this model \citep[see Section 4 in][]{Metzger19} is that there should be a broadband (incoherent) synchrotron afterglow that will accompany and follow the FRB. On timescales similar to the FRB duration, this afterglow will peak in hard X-rays/gamma-rays, but it can subsequently cascade through optical bands on timescales of minutes post-burst.  Assuming a plasmoid ejection event with energy $E_{\mathrm{flare}}$ that lasts for a duration $\delta t$, and a fractional magnetization $\sigma$ in the material upstream from the forward shock, the peak frequency $\nu_{\mathrm{syn}}$ of this synchrotron afterglow will vary with time $t$ from the burst event approximately as \citep[following equations (56)-(57) in][]{Metzger19}

\begin{eqnarray}
    h \nu_{\mathrm{syn}}(t_{\mathrm{dec}}) & = & 57~\mathrm{MeV} \left(\frac{\sigma}{0.1}\right)^{1/2} \left(\frac{E}{10^{43}~\mathrm{erg}} \right)^{1/2} \left(\frac{\delta t}{10^{-3}~\mathrm{s}}\right)^{-3/2} \\
    h \nu_{\mathrm{syn}} & = & \begin{cases} h \nu_{\mathrm{syn}}(t_{\mathrm{dec}}) \left(\frac{t}{t_{\mathrm{dec}}}\right)^{-1}, & t < t_{\mathrm{dec}} \\
    h \nu_{\mathrm{syn}}(t_{\mathrm{dec}}) \left(\frac{t}{t_{\mathrm{dec}}}\right)^{-3/2}, & t > t_{\mathrm{dec}}
    \end{cases}
\end{eqnarray}

\noindent where $t_{\mathrm{dec}} \approx \delta t \approx 10^{-4}~\text{s}$, which is a fiducial parameter and can be set to the observed duration of the burst.  The peak synchrotron frequency cascades down to the synchrotron cooling frequency ($\nu_{c}$), which depends on properties of the circumburst material. Motivated by constraints on the engine of FRB121102 \citep{Margalit18}, \cite{Metzger19} considered previously ejected baryonic-shells as the circumburst material, and parameterized $\nu_{c}$ in terms of the velocity of the ejected material ($\beta=v/c$), the average rate at which this material is injected into the surrounding medium ($\dot{M}$), and the characteristic time between ejection events ($\Delta T$). Following equation (60) in \citet{Metzger19}, the cooling frequency is

\begin{multline}
    h \nu_{c} = 9~\mathrm{keV} \left(\frac{\sigma}{0.1}\right)^{-3/2} \left(\frac{\beta}{0.5} \right)^{3} \left(\frac{\dot{M}}{10^{21}~\text{g s}^{-1}} \right)^{-1} \\
    \left(\frac{t}{10^{-3}~\text{s}} \right)^{-1/2}
    \left(\frac{\Delta T}{10^{5}~\text{s}} \right)^{2}.\label{eqn:cooling}
\end{multline}

Critically, the average period $\Delta T$ between bursts, although directly observable and constrained for FRB\,180916 as 16.3~days \citep{CHIME20}, simply depends on the circumburst density under the assumption that the surrounding medium is filled by ions from previous mass ejection events.  In \autoref{eqn:cooling}, we assume that this medium is characterized by a series of discrete ion shells with a number density of ions in the surrounding medium $n_{\mathrm{ext}}\propto(\Delta T)^{-2}$ where the density profile with radius $r$ from the source of the burst is $n_{\mathrm{ext}} \propto r^{-k}$.  For the discrete shells case, we adopt a $k=0$ following the prescription in \citet{Margalit20a}.

We find that this circumburst density profile provides a natural model for the environment of a source with episodic ejections, but this profile need not be the case if the FRB progenitor erupts inside of a steady wind ($k=2$), a low-density, ambient medium, or homologously expanding shells of ejecta from a supernova.  These density profiles would imply significantly different optical evolution than predicted here, which in general is less constraining as the circumburst density at the radius of the shock drops significantly on timescales much longer than the burst duration.  In addition, although the radio burst itself would probe the density of the circumburst medium out to a radius $r_{\mathrm{dec}} \approx 2 \Gamma^{2} c t_{\mathrm{dec}}$ \citep[$\approx10^{12}$--$10^{13}$~cm for $t_{\mathrm{dec}}=10^{-3}$~s and a Lorentz factor $\Gamma$;][]{Metzger19,Margalit20a}, we assume that this density profile continues outward for several decades in distance because most of the optical emission is produced at time $t_{\rm syn} \gg t_{\rm dec}$ when the FRB is emitted (see \autoref{eq:t_syn} below).  This assumption may hold true if the medium is filled via the continuous ejection of material from a magnetar, but it is a caveat to the following analysis.

Throughout the rest of this paper, we transform $\Delta T$ in (i.e., in \autoref{eqn:cooling}) to $n_{\mathrm{ext}}$, representing the circumburst density at a radius from the progenitor $r_{\mathrm{dec}}=2\Gamma^2 c t_{\mathrm{dec}}$, following equation (32) in \citet{Metzger19} such that

\begin{multline}
   h \nu_{c} = 2.3~\mathrm{keV} \left(\frac{\sigma}{0.1}\right)^{-3/2} \left(\frac{n_{\mathrm{ext}}}{10^{3}~\mathrm{cm}^{-3}}\right)^{-1} \left(\frac{t}{10^{-3}~\mathrm{s}}\right)^{-1/2}.
\end{multline}

The optical light curve is luminous until after the time when the peak synchrotron frequency is equal to the cooling frequency, at which point cooling is no longer efficient.  Although the light curve will briefly continue to rise at $t^{1/2}$ in this regime, as $\nu_{\mathrm{syn}}$ drops below the observing frequency ($\approx$3.6--7.5$\times$10$^{14}$~Hz for $gri$) the optical luminosity will decline exponentially.  We include this cutoff in our light curves by rescaling the optical luminosity by $\exp(-(\nu/\nu_{\mathrm{syn}}-1))$ when $\nu > \nu_{\mathrm{syn}}$.  Overall, we use equations (63)-(64) in \citet{Metzger19} to model the peak luminosity ($L_{\mathrm{pk}}$) and specific luminosity ($L_{\nu}$) of the optical light curve at $t>t_{\rm dec}$ as

\begin{eqnarray}
    L_{\mathrm{pk}} & = & 10^{45}~\text{erg s}^{-1} \left(\frac{E_{\mathrm{flare}}}{10^{43}~\text{erg}}\right) \left(\frac{t}{10^{-3}~\text{s}}\right)^{-1} \\
    \nu L_{\nu} & = & \begin{cases} L_{\mathrm{pk}} \left(\frac{\nu}{\nu_{c}}\right)^{4/3} \left(\frac{\nu_{c}}{\nu_{syn}}\right)^{1/2},  & \nu < \nu_{\mathrm{c}} \\
    L_{\mathrm{pk}} \left(\frac{\nu}{\nu_{\mathrm{syn}}}\right)^{1/2},  & \nu_{c} < \nu < \nu_{\mathrm{syn}}.
    \end{cases}
\end{eqnarray}

We note that in \autoref{fig:models} the light curve begins to decline when $\nu=\nu_{\mathrm{syn}}$, which occurs at

\begin{equation}
\label{eq:t_syn}
    t_{\mathrm{syn}} = 82.6~\mathrm{s} \left(\frac{\lambda}{5000~\mathrm{\AA}}\right)^{2/3} \left(\frac{\sigma}{0.1}\right)^{1/3} \left(\frac{E_{\mathrm{flare}}}{10^{43}~\text{erg}}\right)^{1/3}
\end{equation}

\noindent where $\lambda$ is the observed wavelength.  Thus for $r$-band ($\approx$6231~\AA), the timescale for $E_{\mathrm{flare}}=10^{45}$~erg and $\sigma=0.3$ \citep[as in][]{Metzger19} is $\approx$640~s, or about $1/3$ of our observation window.  This is also significantly longer than any individual observation, implying that our limits for the full set or some subset of our observations are much more constraining than limits from individual exposures.

Moreover, the predicted luminosity is comparable to or brighter than our limits in the $10^{45}$~erg case, with the peak occurring around this time at $\nu L_{\nu}=8\times10^{40}$~erg~s$^{-1}$.  We also note that higher densities will result in significantly more luminous bursts with $\nu L_{\nu} \propto n_{\mathrm{ext}}^{5/6}$ as we are always in the regime where the optical frequency is below the cooling frequency.

\begin{figure}
    \centering
    \includegraphics[width=0.49\textwidth]{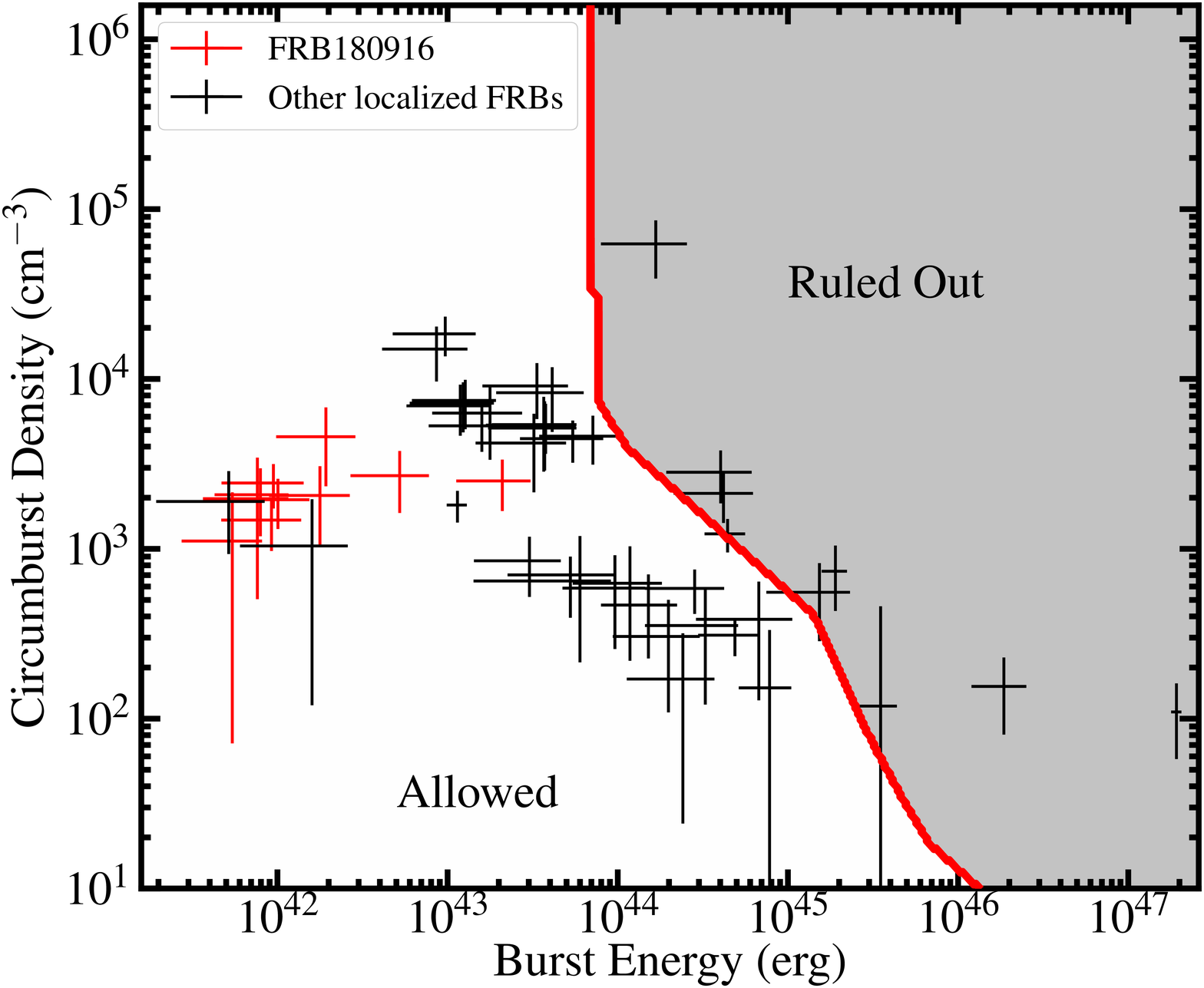}
    \caption{Flare energy ($E_{\mathrm{flare}}$) and circumburst density ($n_{\mathrm{ext}}$) parameter space for FRBs for \citet{Metzger19} synchrotron maser light curves as described in \autoref{sec:analysis}.  We show the range inferred for individual bursts as error bars (from \citealt{Margalit20b}) and FRB\,180916 bursts shown in red.  We demonstrate the energy-density parameter range that we can rule out for the 3 September 2020 burst of FRB\,180916 and using our optical limits as a grey region.}
    \label{fig:parameters}
\end{figure}

Assuming the synchrotron maser model with the same fiducial parameters given above, we consider varying the energy scale of the burst $\log(E_{\mathrm{flare}}/\text{erg}) \in [41.2, 47.4]$ and circumburst ion density $\log(n_{\mathrm{ext}}/\text{cm}^{-3}) \in [1, 6.2]$.  We then model the total in-band emission for our $gri$ observations by calculating the average specific luminosity integrated over the time of each observation relative to the burst arrival time and averaged over frequencies corresponding to the filter response function.  We convert this value to a predicted apparent magnitude assuming the distance modulus and foreground extinction given above.  For each model in our grid, if the computed apparent magnitude is brighter than any of our 3$\sigma$ limiting magnitudes given in \autoref{tab:observations} (including the 3$\sigma$ limits for the stacked, post-burst imaging), we consider that model ruled out as shown through the greyed-out region from \autoref{fig:parameters}.

This model assumes that both the circumburst and interstellar host extinction are negligible in the context of likely optical counterparts.  In an ion-rich medium the optical depth would be dominated by electron scattering with $\tau \approx 10^{-7}$ following equation (33) in \citet{Metzger19}.

The interstellar burst host extinction could be dominated by dust with no indication in the radio signal, and indeed, FRB\,180916 appears coincident with a small enhancement in the optical emission from its host galaxy \citep{Marcote20,Tendulkar20}.  This finding suggests that FRB\,180916 is located within or near a star-forming region \citep[it is 250~pc from a young stellar clump based on a H$\alpha$ detection in {\it Hubble Space Telescope} imaging;][]{Tendulkar20} where there could be excess gas and dust obscuring the optical counterpart \citep[similar to, e.g., stripped-envelope SNe, which evolve from very massive, young stars and thus are observed with high average interstellar host extinction near their birth environments;][]{Stritzinger17}.

A constraint on the total extinction in the FRB\,180916 host galaxy comes from the DM observed toward this source, which is known to be on average $\approx$350~pc~cm$^{-3}$ and 352.6$\pm$3.2~pc~cm$^{-3}$ for the 3 September 2020 burst.  From this value, we account for Milky Way interstellar dispersion adopting 171.7~pc~cm$^{-3}$ on the line-of-sight toward FRB\,180916 \citep[using the NE2001 model of][]{Cordes02} as well as 50~pc~cm$^{-3}$ for the Milky Way halo, although this latter value ranges from 30--80~pc~cm$^{-3}$ \citep[e.g.][]{Prochaska19,Platts20}.  To account for the intergalactic DM, we adopt the relation in \citet{Macquart20} at $z=0.0337$, which gives $56\pm20$~pc~cm$^{-3}$.  Multiplying the residual DM by $1+z$, we obtain a source-frame DM host-galaxy contribution in the line-of-sight to FRB\,180916 of $75\pm30$~pc~cm$^{-3}$. This is equivalent to a hydrogen column density of $N_{H}=2.3\pm1.5\times10^{21}$~cm$^{-2}$ following the locally-derived relation in \citet{He13}.  Finally, this column yields $A_{V}\approx1.0\pm0.6$~mag following \citet{Guver09}, but we acknowledge a significant systematic uncertainty on this value.  Fitting the dust content in the FRB\,180916 host yields a much lower dust content of $E(B-V)=0.12$~mag \citep{Heintz20}, which implies $A_{V}=0.4$~mag assuming $R_{V}=3.1$.  Although this value is not a line-of-sight probe similar to the DM, it is nominally consistent with the lower bound of our interstellar host extinction estimate.  Therefore, we conservatively adopt $A_{V}=0.50$~mag with $R_{V}=3.1$ to model our observations below and assume a \citet{Cardelli89} reddening relation, implying that $A_{g}=0.53$~mag, $A_{r}=0.37$~mag, and $A_{i}=0.27$~mag due to interstellar dust in the host of FRB\,180916.

To place our final limiting magnitudes in context, we consider the full range of circumburst densities corresponding to FRB\,180916 in \citet{Margalit20a}.  In general, the energy of the burst $E_{\mathrm{flare}}$ can be estimated from the equivalent isotropic energy in a single radio burst ($E_{\mathrm{radio}}$) following \citet{Margalit20b,Margalit20a} as

\begin{equation}
    \frac{E_{\mathrm{radio}}}{E_{\mathrm{flare}}} \approx 8.6\times10^{-3} f_{e} f_{\xi} \left(\frac{\nu_{\mathrm{obs}}~t_\mathrm{FRB}}{1~\mathrm{GHz~ms}}\right)^{-1/5}
\end{equation}

\noindent where the observation frequency $\nu_{\mathrm{obs}}$ and the total burst duration $t_{\mathrm{FRB}}$ are known from radio observations, the ratio of electron to ion number densities in the upstream medium $f_{e}=0.5$ is assumed, and the synchrotron maser efficiency is $f_{\xi}\approx10^{-3}$ \citep{Plotnikov19}.  We do not currently know the fluence ($S_{\nu}$) or duration of the 3 September 2020 radio burst from FRB\,180916, and so we assume that it followed the distribution from \citet{CHIME20}, with $E_{\mathrm{radio}}=4\pi D^{2} \nu S_{\nu}$ at 400~MHz of (1.1--39.8)$\times$10$^{37}$~erg and $t_{\mathrm{FRB}}$ of $\approx$0.6--8.6~ms (corresponding to $E_{\mathrm{flare}}=0.04$--1.45$\times10^{43}$~erg).   Assuming an average $E_{\mathrm{radio}}=6.8\times10^{37}$~erg and $t_{\mathrm{FRB}}=3.7$~ms for bursts with well-measured fluence and duration in \citet{CHIME20}, we estimate that the average energy per burst is $E_{\mathrm{flare}}=2.5\times10^{42}$~erg for FRB\,180916 as shown in \autoref{fig:parameters}.

This places the average optical counterpart well outside the range of detectability for our observations and the model parameters above \citep[noting the greyed-out region in \autoref{fig:parameters} and assuming FRB\,180916 parameters -red error bars- from][]{Margalit20a}.  From the expected moment of dispersion-corrected burst arrival at optical wavelengths, the timescale for an optical light curve with $E_{\mathrm{flare}}=2.5\times10^{42}$~erg and $\sigma=0.3$ is $t_{\mathrm{syn}}\approx87$~s, implying that most of our imaging after this point is not very constraining in the context of likely optical counterparts.

Moreover, for a circumburst density $n_{\mathrm{ext}}=2000$~cm$^{-3}$ \citep[roughly the median value for FRB\,180916 in][]{Margalit20a}, the burst would have an optical luminosity $\nu L_{\nu}\approx6\times10^{38}$~erg~s$^{-1}$ ($M$$\approx$$-8$~mag or $m$$\approx$$28$~mag at 150~Mpc) on the timescale $t_{\mathrm{syn}}$.  This is below the threshold of detectability for nearly all optical telescopes, even assuming infinite integration time.  For high-speed optical imagers that can observe on the timescale of tens of ms around a burst \citep[e.g., the 2.4m Thai National Telescope/ULTRASPEC with $m_{\mathrm{limit}}$$\approx$$16.8$~mag over 70~ms;][]{Hardy17}, the detection threshold is shallower by several orders of magnitude.  Thus to detect a burst at optical wavelengths with a light curve similar to those above, a large aperture telescope and an anomalously energetic burst with $E_{\mathrm{flare}}>10^{44}$~erg would be needed (following our limits and the maximum densities inferred for FRB\,180916 in \autoref{fig:parameters}).

On the other hand, if the source were in a highly active state with significantly larger energies and shorter timescale between bursts, this might boost the circumburst density and the corresponding optical signal and potentially place the counterpart within the range of detectability.  The shortest timescales between bursts for FRB\,180916 are only  0.5 milliseconds \citep[observed on 19 Dec 2019;][]{CHIME20}, or $\approx$100~s in cases where $\Delta T/t_{\mathrm{FRB}}\gg1$.  Taking $\Delta T$=100--1.4$\times$10$^{6}$~s as the full range of burst periods and assuming an extremely energetic burst with $E_{\mathrm{flare}}=$10$^{44}$~erg, the circumburst density might exceed $10^{4}$~cm$^{-3}$ \citep[as with FRB\,121102, the burst with the highest inferred circumburst density in][]{Margalit20a}.  Based on the assumed $n_{\mathrm{ext}}\propto\Delta T^{-2}$ and $L_{\mathrm{pk}}\propto n_{\mathrm{ext}}^{5/6}$ scaling above, this would require a change in average period of at least a factor of 5 or $\Delta T\approx3.3$~days.

The FRB may have outbursts on timescales significantly shorter than 3.3~days \citep{CHIME20}, although it is unclear whether these bursts would have similar flare energies or lead to a significantly denser circumburst medium in the magnetar-driven synchrotron maser model we adopt above.  Moreover, the $k=0$ density profile discussed above may not be representative at all projected radii if the burst properties change significantly with time.

Thus while it is unclear if FRB\,180916 can briefly enter the parameter range we rule out in \autoref{fig:parameters}, we would only require moderately deeper imaging or more frequent and energetic bursts to detect the optical counterpart.  Finally, detailed analysis of the 3 September 2020 radio burst detected by CHIME will provide a direct constraint on the circumburst density \citep[following equation (8) in][]{Margalit20a}.

For future optical follow up efforts, this implies that in the context of the baryonic-shell version of the synchrotron maser model, the most promising search strategies will be to target FRB\,180916 with a 8--10~m telescopes when it is in an active state and likely to have one or more bursts.  On the other hand, if the burst profile is significantly more luminous on short timescales after the burst \citep[e.g., following the models of][]{Beloborodov20}, high-speed cameras such as ULTRASPEC \citep{Hardy17} on large aperture telescopes will yield the strongest constraints on potential optical counterparts.  Moreover, if the burst period decreases and the circumburst density is temporarily enhanced, the optical counterpart may be bright such that prompt and/or long-lived counterparts are detectable.

\section{Conclusions}\label{sec:conclusions}

We presented APO 3.5m/ARCTIC observations of FRB\,180916 around the time of a fast radio burst.  Comparing to models of synchrotron maser emission corresponding to the broadband, relatively long-lived counterpart to a radio burst, we find:

\begin{enumerate}
    \item Our observations are constraining for synchrotron maser emission in cases where the energy per burst is larger than $\approx10^{44}$~erg and the circumburst density is greater than $10^{4}$~cm$^{-3}$.
    \item Comparing to previous constraints on the FRB\,180916 burst energy from \citet{Margalit20b}, our limits are not constraining for the predicted burst parameters.  However, if the circumburst density is temporarily enhanced when the FRB progenitor is highly active and multiple discrete bursts occur \citep[e.g., on 19 Dec 2019 and 4 Feb 2020;][]{CHIME20}, the predicted optical light curve could exceed the magnitude limit achievable by large aperture telescopes.
    \item Future optical follow up efforts that target emission similar to the synchrotron maser model will benefit from joint optical and radio observations of FRB\,180916 longer than $30$~min and during its active phase when multiple bursts are likely to occur.
\end{enumerate}

\acknowledgments

We thank J. X. Prochaska for helpful comments on this manuscript.
C.D.K. acknowledges support through a NASA grant in support of {\it Hubble Space Telescope} program AR-16136.
K.E.H. acknowledges support by a Project Grant (162948--051) from The Icelandic Research Fund.
W.F. acknowledges support by the National Science Foundation under grant Nos. AST-1814782 and AST-1909358.
B.M. is supported by NASA through the NASA Hubble Fellowship grant \#HST-HF2-51412.001-A awarded by the Space Telescope Science Institute, which is operated by the Association of Universities for Research in Astronomy, Inc., for NASA, under contract NAS5-26555.
Based on observations obtained with the Apache Point Observatory 3.5-meter telescope, which is owned and operated by the Astrophysical Research Consortium.

\begin{deluxetable*}
{ccccccccc}
\tabletypesize{\scriptsize}
\tablecaption{APO 3.5m/ARCTIC Observations of FRB\,180916\label{tab:observations}}
\tablewidth{0pt}
\tableheadfrac{0.15}
\tablehead{
\colhead{Shutter Open (Epoch)\tablenotemark{a}} &
\colhead{Shutter Closed (Epoch)\tablenotemark{a}} &
\colhead{Cumulative Exposure} &
\colhead{Filter} &
\colhead{$\alpha$\tablenotemark{b}} &
\colhead{$\delta$\tablenotemark{b}} &
\colhead{$m_{\mathrm{lim}}$\tablenotemark{c}} &
\colhead{$M_{\mathrm{lim}}$\tablenotemark{d}} \\
(s) &
(s) &
(s) &
&
(J2000) &
(J2000) &
(AB mag) &
(AB mag)}
\startdata
11:05:39.7503Z (-283.6) & 11:06:10.0854Z (-253.3) & 30.335 & $i$ & 01:57:58.831 & +65:43:05.13 & 24.87 & $-$12.77 \\
11:07:05.7701Z (-197.6) & 11:07:36.1048Z (-167.3) & 30.335 & $i$ & 01:57:57.938 & +65:43:05.02 & 24.75 & $-$12.88 \\
11:08:20.3947Z (-123.0) & 11:08:50.7298Z (-92.7) & 30.335 & $i$ & 01:57:57.972 & +65:43:00.04 & 24.82 & $-$12.81 \\
11:09:27.4379Z (-56.0) & 11:09:57.7672Z (-25.6) & 30.329 & $r$ & 01:57:58.777 & +65:43:05.05 & 24.86 & $-$13.36 \\
11:10:25.5880Z (2.2) & 11:10:55.9377Z ( 32.5) & 30.350 & $r$ & 01:57:57.966 & +65:43:05.14 & 24.87 & $-$13.35 \\
11:11:31.1064Z (67.7) & 11:12:01.4421Z ( 98.0) & 30.336 & $r$ & 01:57:57.937 & +65:42:59.95 & 25.02 & $-$13.21 \\
11:12:43.4415Z (140.0) & 11:13:13.7769Z (170.4) & 30.335 & $g$ & 01:57:58.739 & +65:43:05.26 & 24.60 & $-$14.65 \\
11:13:41.9013Z (198.5) & 11:14:12.2416Z (228.8) & 30.340 & $g$ & 01:57:57.941 & +65:43:05.20 & 24.62 & $-$14.63 \\
11:14:39.9834Z (256.6) & 11:15:10.3232Z (286.9) & 30.340 & $g$ & 01:57:57.935 & +65:43:00.17 & 24.42 & $-$14.83 \\
11:15:52.9075Z (329.5) & 11:16:23.2385Z (359.8) & 30.331 & $i$ & 01:57:58.825 & +65:43:04.73 & 25.00 & $-$12.64 \\
11:16:51.1111Z (387.7) & 11:17:21.4439Z (418.0) & 30.333 & $i$ & 01:57:57.990 & +65:43:04.33 & 24.94 & $-$12.69 \\
11:17:50.0614Z (446.7) & 11:18:20.4010Z (477.0) & 30.340 & $i$ & 01:57:57.972 & +65:42:59.66 & 24.79 & $-$12.84 \\
11:18:56.7009Z (513.3) & 11:19:27.0366Z (543.6) & 30.336 & $r$ & 01:57:58.771 & +65:43:04.94 & 24.85 & $-$13.38 \\
11:19:55.0051Z (571.6) & 11:20:25.3396Z (601.9) & 30.334 & $r$ & 01:57:57.993 & +65:43:05.02 & 24.80 & $-$13.43 \\
11:20:53.2300Z (629.8) & 11:21:23.5680Z (660.2) & 30.338 & $r$ & 01:57:57.947 & +65:42:59.99 & 24.96 & $-$13.27 \\
11:22:03.6474Z (700.2) & 11:22:33.9856Z (730.6) & 30.338 & $g$ & 01:57:58.721 & +65:43:05.29 & 24.58 & $-$14.67 \\
11:23:01.7765Z (758.4) & 11:23:32.1159Z (788.7) & 30.339 & $g$ & 01:57:57.961 & +65:43:05.07 & 24.51 & $-$14.75 \\
11:23:59.3999Z (816.0) & 11:24:29.7394Z (846.3) & 30.339 & $g$ & 01:57:57.981 & +65:43:00.28 & 24.56 & $-$14.69 \\
11:25:12.1161Z (888.7) & 11:25:42.4525Z (919.1) & 30.336 & $i$ & 01:57:58.801 & +65:43:04.87 & 25.02 & $-$12.62 \\
11:26:10.6413Z (947.2) & 11:26:40.9768Z (977.6) & 30.335 & $i$ & 01:57:57.963 & +65:43:04.80 & 25.02 & $-$12.61 \\
11:27:08.5896Z (1005.2) & 11:27:38.8951Z (1035.5) & 30.306 & $i$ & 01:57:57.980 & +65:42:59.81 & 24.89 & $-$12.75 \\
11:28:16.3330Z (1072.9) & 11:28:46.6618Z (1103.3) & 30.329 & $r$ & 01:57:58.774 & +65:43:04.93 & 25.03 & $-$13.20 \\
11:29:13.7910Z (1130.4) & 11:29:44.1262Z (1160.7) & 30.335 & $r$ & 01:57:57.969 & +65:43:04.86 & 25.09 & $-$13.14 \\
11:30:12.4674Z (1189.1) & 11:30:42.8033Z (1219.4) & 30.336 & $r$ & 01:57:57.988 & +65:43:00.04 & 25.08 & $-$13.15 \\
11:31:20.9344Z (1257.5) & 11:31:51.2694Z (1287.9) & 30.335 & $g$ & 01:57:58.758 & +65:43:05.19 & 24.64 & $-$14.61 \\
11:32:20.6130Z (1317.2) & 11:32:50.9150Z (1347.5) & 30.302 & $g$ & 01:57:57.943 & +65:43:05.17 & 24.65 & $-$14.60 \\
11:33:18.7291Z (1375.3) & 11:33:49.0651Z (1405.7) & 30.336 & $i$ & 01:57:57.969 & +65:43:00.09 & 22.73 & $-$14.90 \\
11:34:16.2548Z (1432.9) & 11:34:46.5853Z (1463.2) & 30.331 & $i$ & 01:57:58.808 & +65:43:04.75 & 24.77 & $-$12.86 \\
11:35:13.6288Z (1490.2) & 11:35:43.9768Z (1520.6) & 30.348 & $i$ & 01:57:57.987 & +65:43:04.90 & 24.83 & $-$12.80 \\
11:36:25.2157Z (1561.8) & 11:36:55.5512Z (1592.2) & 30.335 & $r$ & 01:57:57.957 & +65:42:59.96 & 25.57 & $-$12.66 \\
11:37:23.8992Z (1620.5) & 11:37:54.1358Z (1650.7) & 30.237 & $r$ & 01:57:58.743 & +65:43:05.36 & 24.97 & $-$13.26 \\
11:38:21.0860Z (1677.7) & 11:38:51.4219Z (1708.0) & 30.336 & $r$ & 01:57:57.974 & +65:43:04.88 & 24.88 & $-$13.35 \\
11:39:19.1340Z (1735.7) & 11:39:49.4752Z (1766.1) & 30.341 & $g$ & 01:57:58.003 & +65:43:00.10 & 25.99 & $-$13.26 \\
11:40:16.5898Z (1793.2) & 11:40:46.9242Z (1823.5) & 30.334 & $g$ & 01:57:58.753 & +65:43:05.38 & 24.40 & $-$14.86 \\
11:41:13.7692Z (1850.4) & 11:41:44.1046Z (1880.7) & 30.335 & $g$ & 01:57:57.950 & +65:43:05.31 & 24.45 & $-$14.81 \\
11:42:11.2244Z (1907.8) & 11:42:41.5377Z (1938.1) & 30.313 & $g$ & 01:57:57.934 & +65:43:05.21 & 24.39 & $-$14.86 \\ \hline
11:10:25.5880Z (2.2) & 11:38:51.4219Z (1708.0) & 364.10\tablenotemark{e} & $r$ & 01:57:57.877 & +65:43:05.08 & 25.83 & $-$12.40 \\
11:12:43.4415Z (140.0) & 11:42:41.5377Z (1938.1) & 364.10\tablenotemark{e} & $g$ & 01:57:57.877 & +65:43:05.08 & 25.84 & $-$13.41 \\
11:15:52.9075Z (329.5) & 11:35:43.9768Z (1520.6) & 273.08\tablenotemark{e} & $i$ & 01:57:57.877 & +65:43:05.08 & 25.70 & $-$11.94 \\
\enddata
\tablenotetext{a}{All times are UTC on 2020-09-03.  The relative epoch is given in seconds compared with the dispersion-corrected burst arrival time (UTC 2020-09-03 11:10:23.4) as described in \autoref{sec:obs}.}
\tablenotetext{b}{Pointing center of our ARCTIC observation.  Note that ARCTIC is a 2048$\times$2048 imager with $\approx$0.232\arcsec\ pixels, for a 7.92\arcmin$\times$7.92\arcmin\ field of view.}
\tablenotetext{c}{3$\sigma$ apparent limiting magnitude at the location of FRB\,180916 averaged over the entire exposure.}
\tablenotetext{d}{3$\sigma$ absolute limiting magnitude accounting for a distance modulus $\mu=35.92$~mag (\autoref{sec:analysis}) and foreground extinction from \citet{Schlafly11}.  The values in this table assume no interstellar host extinction, but we adopt $A_{g}=0.53$~mag, $A_{r}=0.37$~mag, and $A_{i}=0.27$~mag following the discussion in \autoref{sec:analysis}.}
\tablenotetext{e}{Stacked exposure for all imaging after the expected optical arrival time of the 3 September 2020 burst for FRB\,180916.  The limiting magnitude is 3$\sigma$ calculated empirically in the stacked frame as described in \autoref{sec:analysis}.}
\end{deluxetable*}

\textit{Facilities}: APO 3.5m (ARCTIC)

\bibliography{frb}

\begin{thebibliography}{}
\expandafter\ifx\csname natexlab\endcsname\relax\def\natexlab#1{#1}\fi

\bibitem[{{Ai} {et~al.}(2020){Ai}, {Gao}, \& {Zhang}}]{Ai20}
{Ai}, S., {Gao}, H., \& {Zhang}, B. 2020, arXiv e-prints, arXiv:2007.02400

\bibitem[{{Andreoni} {et~al.}(2020){Andreoni}, {Lu}, {Smith}, {Masci}, {Bellm},
  {Graham}, {Kaplan}, {Kasliwal}, {Kaye}, {Kupfer}, {Laher}, {Mahabal},
  {Nordin}, {Porter}, {Prince}, {Reiley}, {Riddle}, {Van Roestel}, \&
  {Yao}}]{Andreoni20}
{Andreoni}, I., {Lu}, W., {Smith}, R.~M., {et~al.} 2020, \apjl, 896, L2

\bibitem[{{Bannister} {et~al.}(2019){Bannister}, {Deller}, {Phillips},
  {Macquart}, {Prochaska}, {Tejos}, {Ryder}, {Sadler}, {Shannon}, {Simha},
  {Day}, {McQuinn}, {North-Hickey}, {Bhandari}, {Arcus}, {Bennert}, {Burchett},
  {Bouwhuis}, {Dodson}, {Ekers}, {Farah}, {Flynn}, {James}, {Kerr}, {Lenc},
  {Mahony}, {O'Meara}, {Os{\l}owski}, {Qiu}, {Treu}, {U}, {Bateman}, {Bock},
  {Bolton}, {Brown}, {Bunton}, {Chippendale}, {Cooray}, {Cornwell}, {Gupta},
  {Hayman}, {Kesteven}, {Koribalski}, {MacLeod}, {McClure-Griffiths},
  {Neuhold}, {Norris}, {Pilawa}, {Qiao}, {Reynolds}, {Roxby}, {Shimwell},
  {Voronkov}, \& {Wilson}}]{Bannister19}
{Bannister}, K.~W., {Deller}, A.~T., {Phillips}, C., {et~al.} 2019, Science,
  365, 565

\bibitem[{{Becker}(2015)}]{hotpants}
{Becker}, A. 2015, {HOTPANTS: High Order Transform of PSF ANd Template
  Subtraction}, ascl:1504.004

\bibitem[{{Beloborodov}(2017)}]{Beloborodov17}
{Beloborodov}, A.~M. 2017, \apjl, 843, L26

\bibitem[{{Beloborodov}(2020)}]{Beloborodov20}
---. 2020, \apj, 896, 142

\bibitem[{{Beniamini} {et~al.}(2020){Beniamini}, {Wadiasingh}, \&
  {Metzger}}]{Beniamini20}
{Beniamini}, P., {Wadiasingh}, Z., \& {Metzger}, B.~D. 2020, \mnras, 496, 3390

\bibitem[{{Best} \& {Bazo}(2019)}]{Best19}
{Best}, S., \& {Bazo}, J. 2019, \jcap, 2019, 004

\bibitem[{{Bhandari} {et~al.}(2018){Bhandari}, {Keane}, {Barr}, {Jameson},
  {Petroff}, {Johnston}, {Bailes}, {Bhat}, {Burgay}, {Burke-Spolaor}, {Caleb},
  {Eatough}, {Flynn}, {Green}, {Jankowski}, {Kramer}, {Krishnan}, {Morello},
  {Possenti}, {Stappers}, {Tiburzi}, {van Straten}, {Andreoni}, {Butterley},
  {Chand ra}, {Cooke}, {Corongiu}, {Coward}, {Dhillon}, {Dodson}, {Hardy},
  {Howell}, {Jaroenjittichai}, {Klotz}, {Littlefair}, {Marsh}, {Mickaliger},
  {Muxlow}, {Perrodin}, {Pritchard}, {Sawangwit}, {Terai}, {Tominaga}, {Torne},
  {Totani}, {Trois}, {Turpin}, {Niino}, {Wilson}, {Albert}, {Andr{\'e}},
  {Anghinolfi}, {Anton}, {Ardid}, {Aubert}, {Avgitas}, {Baret},
  {Barrios-Mart{\'\i}}, {Basa}, {Belhorma}, {Bertin}, {Biagi}, {Bormuth},
  {Bourret}, {Bouwhuis}, {Br{\^a}nza{\textcommabelow s}}, {Bruijn}, {Brunner},
  {Busto}, {Capone}, {Caramete}, {Carr}, {Celli}, {Moursli}, {Chiarusi},
  {Circella}, {Coelho}, {Coleiro}, {Coniglione}, {Costantini}, {Coyle},
  {Creusot}, {D{\'\i}az}, {Deschamps}, {De Bonis}, {Distefano}, {Palma},
  {Domi}, {Donzaud}, {Dornic}, {Drouhin}, {Eberl}, {Bojaddaini}, {Khayati},
  {Els{\"a}sser}, {Enzenh{\"o}fer}, {Ettahiri}, {Fassi}, {Felis}, {Fusco},
  {Gay}, {Giordano}, {Glotin}, {Gregoire}, {Gracia-Ruiz}, {Graf}, {Hallmann},
  {van Haren}, {Heijboer}, {Hello}, {Hern{\'a}ndez-Rey}, {H{\"o}{\ss}l},
  {Hofest{\"a}dt}, {Hugon}, {Illuminati}, {James}, {de Jong}, {Jongen},
  {Kadler}, {Kalekin}, {Katz}, {Kie{\ss}ling}, {Kouchner}, {Kreter},
  {Kreykenbohm}, {Kulikovskiy}, {Lachaud}, {Lahmann}, {Lef{\`e}vre}, {Leonora},
  {Loucatos}, {Marcelin}, {Margiotta}, {Marinelli}, {Mart{\'\i}nez-Mora},
  {Mele}, {Melis}, {Michael}, {Migliozzi}, {Moussa}, {Navas}, {Nezri},
  {Organokov}, {P{\v{a}}v{\v{a}}la{\textcommabelow s}}, {Pellegrino},
  {Perrina}, {Piattelli}, {Popa}, {Pradier}, {Quinn}, {Racca}, {Riccobene},
  {S{\'a}nchez-Losa}, {Salda{\~n}a}, {Salvadori}, {Samtleben}, {Sanguineti},
  {Sapienza}, {Sch{\"u}ssler}, {Sieger}, {Spurio}, {Stolarczyk}, {Taiuti},
  {Tayalati}, {Trovato}, {Turpin}, {T{\"o}nnis}, {Vallage}, {Van Elewyck},
  {Versari}, {Vivolo}, {Vizzocca}, {Wilms}, {Zornoza}, \&
  {Z{\'u}{\~n}iga}}]{Bhandari18}
{Bhandari}, S., {Keane}, E.~F., {Barr}, E.~D., {et~al.} 2018, \mnras, 475, 1427

\bibitem[{{Bochenek} {et~al.}(2020){Bochenek}, {Ravi}, \& {Dong}}]{Bochenek20}
{Bochenek}, C.~D., {Ravi}, V., \& {Dong}, D. 2020, arXiv e-prints,
  arXiv:2009.13030

\bibitem[{{Caleb} {et~al.}(2019){Caleb}, {Stappers}, {Rajwade}, \&
  {Flynn}}]{Caleb19}
{Caleb}, M., {Stappers}, B.~W., {Rajwade}, K., \& {Flynn}, C. 2019, \mnras,
  484, 5500

\bibitem[{{Cardelli} {et~al.}(1989){Cardelli}, {Clayton}, \&
  {Mathis}}]{Cardelli89}
{Cardelli}, J.~A., {Clayton}, G.~C., \& {Mathis}, J.~S. 1989, \apj, 345, 245

\bibitem[{{Chatterjee} {et~al.}(2017){Chatterjee}, {Law}, {Wharton},
  {Burke-Spolaor}, {Hessels}, {Bower}, {Cordes}, {Tendulkar}, {Bassa},
  {Demorest}, {Butler}, {Seymour}, {Scholz}, {Abruzzo}, {Bogdanov}, {Kaspi},
  {Keimpema}, {Lazio}, {Marcote}, {McLaughlin}, {Paragi}, {Ransom}, {Rupen},
  {Spitler}, \& {van Langevelde}}]{Chatterjee17}
{Chatterjee}, S., {Law}, C.~J., {Wharton}, R.~S., {et~al.} 2017, \nat, 541, 58

\bibitem[{{Chen} {et~al.}(2020){Chen}, {Ravi}, \& {Lu}}]{Chen20}
{Chen}, G., {Ravi}, V., \& {Lu}, W. 2020, \apj, 897, 146

\bibitem[{{CHIME/FRB Collaboration} {et~al.}(2019){CHIME/FRB Collaboration},
  {Andersen}, {Bandura}, {Bhardwaj}, {Boubel}, {Boyce}, {Boyle}, {Brar},
  {Cassanelli}, {Chawla}, {Cubranic}, {Deng}, {Dobbs}, {Fandino}, {Fonseca},
  {Gaensler}, {Gilbert}, {Giri}, {Good}, {Halpern}, {Hill}, {Hinshaw},
  {H{\"o}fer}, {Josephy}, {Kaspi}, {Kothes}, {Landecker}, {Lang}, {Li}, {Lin},
  {Masui}, {Mena-Parra}, {Merryfield}, {Mckinven}, {Michilli}, {Milutinovic},
  {Naidu}, {Newburgh}, {Ng}, {Patel}, {Pen}, {Pinsonneault-Marotte}, {Pleunis},
  {Rafiei-Ravandi}, {Rahman}, {Ransom}, {Renard}, {Scholz}, {Siegel}, {Singh},
  {Smith}, {Stairs}, {Tendulkar}, {Tretyakov}, {Vanderlinde}, {Yadav}, \&
  {Zwaniga}}]{CHIME19}
{CHIME/FRB Collaboration}, {Andersen}, B.~C., {Bandura}, K., {et~al.} 2019,
  \apjl, 885, L24

\bibitem[{{CHIME/FRB Collaboration} {et~al.}(2020{\natexlab{a}}){CHIME/FRB
  Collaboration}, {Andersen}, {Bandura}, {Bhardwaj}, {Bij}, {Boyce}, {Boyle},
  {Brar}, {Cassanelli}, {Chawla}, {Chen}, {Cliche}, {Cook}, {Cubranic},
  {Curtin}, {Denman}, {Dobbs}, {Dong}, {Fandino}, {Fonseca}, {Gaensler},
  {Giri}, {Good}, {Halpern}, {Hill}, {Hinshaw}, {H{\"o}fer}, {Josephy},
  {Kania}, {Kaspi}, {Landecker}, {Leung}, {Li}, {Lin}, {Masui}, {Mckinven},
  {Mena-Parra}, {Merryfield}, {Meyers}, {Michilli}, {Milutinovic},
  {Mirhosseini}, {M{\"u}nchmeyer}, {Naidu}, {Newburgh}, {Ng}, {Patel}, {Pen},
  {Pinsonneault-Marotte}, {Pleunis}, {Quine}, {Rafiei-Ravandi}, {Rahman},
  {Ransom}, {Renard}, {Sanghavi}, {Scholz}, {Shaw}, {Shin}, {Siegel}, {Singh},
  {Smegal}, {Smith}, {Stairs}, {Tan}, {Tendulkar}, {Tretyakov}, {Vanderlinde},
  {Wang}, {Wulf}, \& {Zwaniga}}]{frb200428}
{CHIME/FRB Collaboration}, {Andersen}, B.~C., {Bandura}, K.~M., {et~al.}
  2020{\natexlab{a}}, arXiv:2005.10324, arXiv:2005.10324

\bibitem[{{CHIME/FRB Collaboration} {et~al.}(2020{\natexlab{b}}){CHIME/FRB
  Collaboration}, {Amiri}, {Andersen}, {Band ura}, {Bhardwaj}, {Boyle}, {Brar},
  {Chawla}, {Chen}, {Cliche}, {Cubranic}, {Deng}, {Denman}, {Dobbs}, {Dong},
  {Fand ino}, {Fonseca}, {Gaensler}, {Giri}, {Good}, {Halpern}, {Hessels},
  {Hill}, {H{\"o}fer}, {Josephy}, {Kania}, {Karuppusamy}, {Kaspi}, {Keimpema},
  {Kirsten}, {Landecker}, {Lang}, {Leung}, {Li}, {Lin}, {Marcote}, {Masui},
  {McKinven}, {Mena-Parra}, {Merryfield}, {Michilli}, {Milutinovic},
  {Mirhosseini}, {Naidu}, {Newburgh}, {Ng}, {Nimmo}, {Paragi}, {Patel}, {Pen},
  {Pinsonneault-Marotte}, {Pleunis}, {Rafiei-Ravandi}, {Rahman}, {Ransom},
  {Renard}, {Sanghavi}, {Scholz}, {Shaw}, {Shin}, {Siegel}, {Singh}, {Smegal},
  {Smith}, {Stairs}, {Tendulkar}, {Tretyakov}, {Vanderlinde}, {Wang}, {Wang},
  {Wulf}, {Yadav}, \& {Zwaniga}}]{CHIME20}
{CHIME/FRB Collaboration}, {Amiri}, M., {Andersen}, B.~C., {et~al.}
  2020{\natexlab{b}}, \nat, 582, 351

\bibitem[{{Cordes} \& {Chatterjee}(2019)}]{Cordes19}
{Cordes}, J.~M., \& {Chatterjee}, S. 2019, \araa, 57, 417

\bibitem[{{Cordes} \& {Lazio}(2002)}]{Cordes02}
{Cordes}, J.~M., \& {Lazio}, T.~J.~W. 2002, arXiv e-prints, astro

\bibitem[{{Cunningham} {et~al.}(2019){Cunningham}, {Cenko}, {Burns},
  {Goldstein}, {Lien}, {Kocevski}, {Briggs}, {Connaughton}, {Miller},
  {Racusin}, \& {Stanbro}}]{Cunningham19}
{Cunningham}, V., {Cenko}, S.~B., {Burns}, E., {et~al.} 2019, \apj, 879, 40

\bibitem[{{DeLaunay} {et~al.}(2016){DeLaunay}, {Fox}, {Murase},
  {M{\'e}sz{\'a}ros}, {Keivani}, {Messick}, {Mostaf{\'a}}, {Oikonomou},
  {Te{\v{s}}i{\'c}}, \& {Turley}}]{Delaunay16}
{DeLaunay}, J.~J., {Fox}, D.~B., {Murase}, K., {et~al.} 2016, \apjl, 832, L1

\bibitem[{{Falcke} \& {Rezzolla}(2014)}]{Falcke14}
{Falcke}, H., \& {Rezzolla}, L. 2014, \aap, 562, A137

\bibitem[{{Flewelling} {et~al.}(2016){Flewelling}, {Magnier}, {Chambers},
  {Heasley}, {Holmberg}, {Huber}, {Sweeney}, {Waters}, {Calamida}, {Casertano},
  {Chen}, {Farrow}, {Hasinger}, {Henderson}, {Long}, {Metcalfe}, {Narayan},
  {Nieto-Santisteban}, {Norberg}, {Rest}, {Saglia}, {Szalay}, {Thakar},
  {Tonry}, {Valenti}, {Werner}, {White}, {Denneau}, {Draper}, {Hodapp},
  {Jedicke}, {Kaiser}, {Kudritzki}, {Price}, {Wainscoat}, {Builders},
  {Chastel}, {McLean}, {Postman}, \& {Shiao}}]{Flewelling16}
{Flewelling}, H.~A., {Magnier}, E.~A., {Chambers}, K.~C., {et~al.} 2016, arXiv
  e-prints, arXiv:1612.05243

\bibitem[{{Gao} \& {Zhang}(2017)}]{Gao17}
{Gao}, H., \& {Zhang}, B. 2017, \apjl, 835, L21

\bibitem[{{Gehrels} {et~al.}(2004){Gehrels}, {Chincarini}, {Giommi}, {Mason},
  {Nousek}, {Wells}, {White}, {Barthelmy}, {Burrows}, {Cominsky}, {Hurley},
  {Marshall}, {M{\'e}sz{\'a}ros}, {Roming}, {Angelini}, {Barbier}, {Belloni},
  {Campana}, {Caraveo}, {Chester}, {Citterio}, {Cline}, {Cropper}, {Cummings},
  {Dean}, {Feigelson}, {Fenimore}, {Frail}, {Fruchter}, {Garmire}, {Gendreau},
  {Ghisellini}, {Greiner}, {Hill}, {Hunsberger}, {Krimm}, {Kulkarni}, {Kumar},
  {Lebrun}, {Lloyd-Ronning}, {Markwardt}, {Mattson}, {Mushotzky}, {Norris},
  {Osborne}, {Paczynski}, {Palmer}, {Park}, {Parsons}, {Paul}, {Rees},
  {Reynolds}, {Rhoads}, {Sasseen}, {Schaefer}, {Short}, {Smale}, {Smith},
  {Stella}, {Tagliaferri}, {Takahashi}, {Tashiro}, {Townsley}, {Tueller},
  {Turner}, {Vietri}, {Voges}, {Ward}, {Willingale}, {Zerbi}, \&
  {Zhang}}]{Gehrels2004}
{Gehrels}, N., {Chincarini}, G., {Giommi}, P., {et~al.} 2004, \apj, 611, 1005

\bibitem[{{Guidorzi} {et~al.}(2019){Guidorzi}, {Marongiu}, {Martone}, {Amati},
  {Frontera}, {Nicastro}, {Orlandini}, {Margutti}, \& {Virgilli}}]{Guidorzi19}
{Guidorzi}, C., {Marongiu}, M., {Martone}, R., {et~al.} 2019, \apj, 882, 100

\bibitem[{{Guidorzi} {et~al.}(2020){Guidorzi}, {Marongiu}, {Martone},
  {Nicastro}, {Xiong}, {Liao}, {Li}, {Zhang}, {Amati}, {Frontera}, {Orlandini},
  {Rosati}, {Virgilli}, {Zhang}, {Bu}, {Cai}, {Cao}, {Chang}, {Chen}, {Chen},
  {Chen}, {Chen}, {Chen}, {Cui}, {Cui}, {Deng}, {Dong}, {Du}, {Fu}, {Gao},
  {Gao}, {Gao}, {Ge}, {Gu}, {Guan}, {Guo}, {Han}, {Huang}, {Huo}, {Jia},
  {Jiang}, {Jiang}, {Jin}, {Jin}, {Kong}, {Li}, {Li}, {Li}, {Li}, {Li}, {Li},
  {Li}, {Li}, {Li}, {Li}, {Liang}, {Liu}, {Liu}, {Liu}, {Liu}, {Liu}, {Liu},
  {Lu}, {Lu}, {Lu}, {Luo}, {Luo}, {Ma}, {Ma}, {Meng}, {Nang}, {Nie}, {Ou},
  {Qu}, {Sai}, {Shang}, {Song}, {Song}, {Sun}, {Tan}, {Tao}, {Tuo}, {Wang},
  {Wang}, {Wang}, {Wang}, {Wang}, {Wen}, {Wu}, {Wu}, {Wu}, {Xiao}, {Xiao},
  {Xu}, {Yang}, {Yang}, {Yang}, {Yi}, {Yin}, {You}, {Zhang}, {Zhang}, {Zhang},
  {Zhang}, {Zhang}, {Zhang}, {Zhang}, {Zhang}, {Zhang}, {Zhang}, {Zhang},
  {Zhang}, {Zhang}, {Zhang}, {Zhang}, {Zhang}, {Zhang}, {Zhang}, {Zheng},
  {Zhou}, {Zhou}, {Zhu}, {Zhu}, \& {Zhuang}}]{Guidorzi20}
---. 2020, \aap, 637, A69

\bibitem[{{G{\"u}ver} \& {{\"O}zel}(2009)}]{Guver09}
{G{\"u}ver}, T., \& {{\"O}zel}, F. 2009, \mnras, 400, 2050

\bibitem[{{Hardy} {et~al.}(2017){Hardy}, {Dhillon}, {Spitler}, {Littlefair},
  {Ashley}, {De Cia}, {Green}, {Jaroenjittichai}, {Keane}, {Kerry}, {Kramer},
  {Malesani}, {Marsh}, {Parsons}, {Possenti}, {Rattanasoon}, \&
  {Sahman}}]{Hardy17}
{Hardy}, L.~K., {Dhillon}, V.~S., {Spitler}, L.~G., {et~al.} 2017, \mnras, 472,
  2800

\bibitem[{{Hashimoto} {et~al.}(2020){Hashimoto}, {Goto}, {On}, {Lu}, {Santos},
  {Ho}, {Wang}, {Kim}, \& {Hsiao}}]{Hashimoto20}
{Hashimoto}, T., {Goto}, T., {On}, A. Y.~L., {et~al.} 2020, \mnras, 497, 4107

\bibitem[{{He} {et~al.}(2013){He}, {Ng}, \& {Kaspi}}]{He13}
{He}, C., {Ng}, C.~Y., \& {Kaspi}, V.~M. 2013, \apj, 768, 64

\bibitem[{{Heintz} {et~al.}(2020){Heintz}, {Prochaska}, {Simha}, {Platts},
  {Fong}, {Tejos}, {Ryder}, {Aggerwal}, {Bhandari}, {Day}, {Deller},
  {Kilpatrick}, {Law}, {Macquart}, {Mannings}, {Marnoch}, {Sadler}, \&
  {Shannon}}]{Heintz20}
{Heintz}, K.~E., {Prochaska}, J.~X., {Simha}, S., {et~al.} 2020, arXiv
  e-prints, arXiv:2009.10747

\bibitem[{{Huehnerhoff} {et~al.}(2016){Huehnerhoff}, {Ketzeback}, {Bradley},
  {Dembicky}, {Doughty}, {Hawley}, {Johnson}, {Klaene}, {Leon}, {McMillan},
  {Owen}, {Sayres}, {Sheen}, \& {Shugart}}]{ARCTIC}
{Huehnerhoff}, J., {Ketzeback}, W., {Bradley}, A., {et~al.} 2016, in Society of
  Photo-Optical Instrumentation Engineers (SPIE) Conference Series, Vol. 9908,
  Ground-based and Airborne Instrumentation for Astronomy VI, 99085H

\bibitem[{{Katz}(2016)}]{Katz16}
{Katz}, J.~I. 2016, \apj, 826, 226

\bibitem[{{Kilpatrick} {et~al.}(2018){Kilpatrick}, {Foley}, {Drout}, {Pan},
  {Panther}, {Coulter}, {Filippenko}, {Marion}, {Piro}, {Rest}, {Seitenzahl},
  {Strampelli}, \& {Wang}}]{kilpatrick+18}
{Kilpatrick}, C.~D., {Foley}, R.~J., {Drout}, M.~R., {et~al.} 2018, \mnras,
  473, 4805

\bibitem[{{Kulkarni} {et~al.}(2014){Kulkarni}, {Ofek}, {Neill}, {Zheng}, \&
  {Juric}}]{Kulkarni14}
{Kulkarni}, S.~R., {Ofek}, E.~O., {Neill}, J.~D., {Zheng}, Z., \& {Juric}, M.
  2014, \apj, 797, 70

\bibitem[{{Kumar} {et~al.}(2017){Kumar}, {Lu}, \& {Bhattacharya}}]{Kumar17}
{Kumar}, P., {Lu}, W., \& {Bhattacharya}, M. 2017, \mnras, 468, 2726

\bibitem[{{Kumar} {et~al.}(2020){Kumar}, {Shannon}, {Flynn}, {Os{\l}owski},
  {Bhandari}, {Day}, {Deller}, {Farah}, {Kaczmarek}, {Kerr}, {Phillips},
  {Price}, {Qiu}, \& {Thyagarajan}}]{Kumar20}
{Kumar}, P., {Shannon}, R.~M., {Flynn}, C., {et~al.} 2020, \mnras,
  arXiv:2009.01214

\bibitem[{{Li} \& {Zhang}(2020)}]{Li20}
{Li}, Y., \& {Zhang}, B. 2020, \apjl, 899, L6

\bibitem[{{Lorimer} {et~al.}(2007){Lorimer}, {Bailes}, {McLaughlin},
  {Narkevic}, \& {Crawford}}]{Lorimer07}
{Lorimer}, D.~R., {Bailes}, M., {McLaughlin}, M.~A., {Narkevic}, D.~J., \&
  {Crawford}, F. 2007, Science, 318, 777

\bibitem[{{Lu} {et~al.}(2020){Lu}, {Kumar}, \& {Zhang}}]{Lu20}
{Lu}, W., {Kumar}, P., \& {Zhang}, B. 2020, \mnras, 498, 1397

\bibitem[{{Lyubarsky}(2014)}]{Lyubarsky14}
{Lyubarsky}, Y. 2014, \mnras, 442, L9

\bibitem[{{Macquart} \& {Ekers}(2018)}]{Macquart18}
{Macquart}, J.~P., \& {Ekers}, R. 2018, \mnras, 480, 4211

\bibitem[{{Macquart} {et~al.}(2020){Macquart}, {Prochaska}, {McQuinn},
  {Bannister}, {Bhandari}, {Day}, {Deller}, {Ekers}, {James}, {Marnoch},
  {Os{\l}owski}, {Phillips}, {Ryder}, {Scott}, {Shannon}, \&
  {Tejos}}]{Macquart20}
{Macquart}, J.~P., {Prochaska}, J.~X., {McQuinn}, M., {et~al.} 2020, \nat, 581,
  391

\bibitem[{{Marcote} {et~al.}(2020){Marcote}, {Nimmo}, {Hessels}, {Tendulkar},
  {Bassa}, {Paragi}, {Keimpema}, {Bhardwaj}, {Karuppusamy}, {Kaspi}, {Law},
  {Michilli}, {Aggarwal}, {Andersen}, {Archibald}, {Bandura}, {Bower}, {Boyle},
  {Brar}, {Burke-Spolaor}, {Butler}, {Cassanelli}, {Chawla}, {Demorest},
  {Dobbs}, {Fonseca}, {Giri}, {Good}, {Gourdji}, {Josephy}, {Kirichenko},
  {Kirsten}, {Landecker}, {Lang}, {Lazio}, {Li}, {Lin}, {Linford}, {Masui},
  {Mena-Parra}, {Naidu}, {Ng}, {Patel}, {Pen}, {Pleunis}, {Rafiei-Ravandi},
  {Rahman}, {Renard}, {Scholz}, {Siegel}, {Smith}, {Stairs}, {Vanderlinde}, \&
  {Zwaniga}}]{Marcote20}
{Marcote}, B., {Nimmo}, K., {Hessels}, J.~W.~T., {et~al.} 2020, \nat, 577, 190

\bibitem[{{Margalit} {et~al.}(2020{\natexlab{a}}){Margalit}, {Beniamini},
  {Sridhar}, \& {Metzger}}]{Margalit20b}
{Margalit}, B., {Beniamini}, P., {Sridhar}, N., \& {Metzger}, B.~D.
  2020{\natexlab{a}}, \apjl, 899, L27

\bibitem[{{Margalit} {et~al.}(2019){Margalit}, {Berger}, \&
  {Metzger}}]{Margalit19b}
{Margalit}, B., {Berger}, E., \& {Metzger}, B.~D. 2019, \apj, 886, 110

\bibitem[{{Margalit} \& {Metzger}(2018)}]{Margalit18}
{Margalit}, B., \& {Metzger}, B.~D. 2018, \apjl, 868, L4

\bibitem[{{Margalit} {et~al.}(2020{\natexlab{b}}){Margalit}, {Metzger}, \&
  {Sironi}}]{Margalit20a}
{Margalit}, B., {Metzger}, B.~D., \& {Sironi}, L. 2020{\natexlab{b}}, \mnras,
  494, 4627

\bibitem[{{Metzger} {et~al.}(2017){Metzger}, {Berger}, \&
  {Margalit}}]{Metzger17}
{Metzger}, B.~D., {Berger}, E., \& {Margalit}, B. 2017, \apj, 841, 14

\bibitem[{{Metzger} {et~al.}(2019){Metzger}, {Margalit}, \&
  {Sironi}}]{Metzger19}
{Metzger}, B.~D., {Margalit}, B., \& {Sironi}, L. 2019, \mnras, 485, 4091

\bibitem[{{Petroff} {et~al.}(2019{\natexlab{a}}){Petroff}, {Hessels}, \&
  {Lorimer}}]{Petroff19}
{Petroff}, E., {Hessels}, J.~W.~T., \& {Lorimer}, D.~R. 2019{\natexlab{a}},
  \aapr, 27, 4

\bibitem[{{Petroff} {et~al.}(2015){Petroff}, {Bailes}, {Barr}, {Barsdell},
  {Bhat}, {Bian}, {Burke-Spolaor}, {Caleb}, {Champion}, {Chandra}, {Da Costa},
  {Delvaux}, {Flynn}, {Gehrels}, {Greiner}, {Jameson}, {Johnston}, {Kasliwal},
  {Keane}, {Keller}, {Kocz}, {Kramer}, {Leloudas}, {Malesani}, {Mulchaey},
  {Ng}, {Ofek}, {Perley}, {Possenti}, {Schmidt}, {Shen}, {Stappers}, {Tisserand
  }, {van Straten}, \& {Wolf}}]{Petroff15}
{Petroff}, E., {Bailes}, M., {Barr}, E.~D., {et~al.} 2015, \mnras, 447, 246

\bibitem[{{Petroff} {et~al.}(2016){Petroff}, {Barr}, {Jameson}, {Keane},
  {Bailes}, {Kramer}, {Morello}, {Tabbara}, \& {van Straten}}]{Petroff16}
{Petroff}, E., {Barr}, E.~D., {Jameson}, A., {et~al.} 2016, \pasa, 33, e045

\bibitem[{{Petroff} {et~al.}(2019{\natexlab{b}}){Petroff}, {Oostrum},
  {Stappers}, {Bailes}, {Barr}, {Bates}, {Bhandari}, {Bhat}, {Burgay},
  {Burke-Spolaor}, {Cameron}, {Champion}, {Eatough}, {Flynn}, {Jameson},
  {Johnston}, {Keane}, {Keith}, {Kramer}, {Levin}, {Morello}, {Ng}, {Possenti},
  {Ravi}, {van Straten}, {Thornton}, \& {Tiburzi}}]{Petroff18}
{Petroff}, E., {Oostrum}, L.~C., {Stappers}, B.~W., {et~al.}
  2019{\natexlab{b}}, \mnras, 482, 3109

\bibitem[{{Pilia} {et~al.}(2020){Pilia}, {Burgay}, {Possenti}, {Ridolfi},
  {Gajjar}, {Corongiu}, {Perrodin}, {Bernardi}, {Naldi}, {Pupillo},
  {Ambrosino}, {Bianchi}, {Burtovoi}, {Casella}, {Casentini}, {Cecconi},
  {Ferrigno}, {Fiori}, {Gendreau}, {Ghedina}, {Naletto}, {Nicastro}, {Ochner},
  {Palazzi}, {Panessa}, {Papitto}, {Pittori}, {Rea}, {Castillo}, {Savchenko},
  {Setti}, {Tavani}, {Trois}, {Trudu}, {Turatto}, {Ursi}, {Verrecchia}, \&
  {Zampieri}}]{Pilia20}
{Pilia}, M., {Burgay}, M., {Possenti}, A., {et~al.} 2020, \apjl, 896, L40

\bibitem[{{Planck Collaboration} {et~al.}(2016){Planck Collaboration}, {Ade},
  {Aghanim}, {Arnaud}, {Ashdown}, {Aumont}, {Baccigalupi}, {Banday},
  {Barreiro}, {Bartlett}, {Bartolo}, {Battaner}, {Battye}, {Benabed},
  {Beno{\^\i}t}, {Benoit-L{\'e}vy}, {Bernard}, {Bersanelli}, {Bielewicz},
  {Bock}, {Bonaldi}, {Bonavera}, {Bond}, {Borrill}, {Bouchet}, {Boulanger},
  {Bucher}, {Burigana}, {Butler}, {Calabrese}, {Cardoso}, {Catalano},
  {Challinor}, {Chamballu}, {Chary}, {Chiang}, {Chluba}, {Christensen},
  {Church}, {Clements}, {Colombi}, {Colombo}, {Combet}, {Coulais}, {Crill},
  {Curto}, {Cuttaia}, {Danese}, {Davies}, {Davis}, {de Bernardis}, {de Rosa},
  {de Zotti}, {Delabrouille}, {D{\'e}sert}, {Di Valentino}, {Dickinson},
  {Diego}, {Dolag}, {Dole}, {Donzelli}, {Dor{\'e}}, {Douspis}, {Ducout},
  {Dunkley}, {Dupac}, {Efstathiou}, {Elsner}, {En{\ss}lin}, {Eriksen},
  {Farhang}, {Fergusson}, {Finelli}, {Forni}, {Frailis}, {Fraisse},
  {Franceschi}, {Frejsel}, {Galeotta}, {Galli}, {Ganga}, {Gauthier}, {Gerbino},
  {Ghosh}, {Giard}, {Giraud-H{\'e}raud}, {Giusarma}, {Gjerl{\o}w},
  {Gonz{\'a}lez-Nuevo}, {G{\'o}rski}, {Gratton}, {Gregorio}, {Gruppuso},
  {Gudmundsson}, {Hamann}, {Hansen}, {Hanson}, {Harrison}, {Helou},
  {Henrot-Versill{\'e}}, {Hern{\'a}ndez-Monteagudo}, {Herranz}, {Hildebrand t},
  {Hivon}, {Hobson}, {Holmes}, {Hornstrup}, {Hovest}, {Huang}, {Huffenberger},
  {Hurier}, {Jaffe}, {Jaffe}, {Jones}, {Juvela}, {Keih{\"a}nen}, {Keskitalo},
  {Kisner}, {Kneissl}, {Knoche}, {Knox}, {Kunz}, {Kurki-Suonio}, {Lagache},
  {L{\"a}hteenm{\"a}ki}, {Lamarre}, {Lasenby}, {Lattanzi}, {Lawrence}, {Leahy},
  {Leonardi}, {Lesgourgues}, {Levrier}, {Lewis}, {Liguori}, {Lilje},
  {Linden-V{\o}rnle}, {L{\'o}pez-Caniego}, {Lubin}, {Mac{\'\i}as-P{\'e}rez},
  {Maggio}, {Maino}, {Mandolesi}, {Mangilli}, {Marchini}, {Maris}, {Martin},
  {Martinelli}, {Mart{\'\i}nez-Gonz{\'a}lez}, {Masi}, {Matarrese}, {McGehee},
  {Meinhold}, {Melchiorri}, {Melin}, {Mendes}, {Mennella}, {Migliaccio},
  {Millea}, {Mitra}, {Miville-Desch{\^e}nes}, {Moneti}, {Montier}, {Morgante},
  {Mortlock}, {Moss}, {Munshi}, {Murphy}, {Naselsky}, {Nati}, {Natoli},
  {Netterfield}, {N{\o}rgaard-Nielsen}, {Noviello}, {Novikov}, {Novikov},
  {Oxborrow}, {Paci}, {Pagano}, {Pajot}, {Paladini}, {Paoletti}, {Partridge},
  {Pasian}, {Patanchon}, {Pearson}, {Perdereau}, {Perotto}, {Perrotta},
  {Pettorino}, {Piacentini}, {Piat}, {Pierpaoli}, {Pietrobon}, {Plaszczynski},
  {Pointecouteau}, {Polenta}, {Popa}, {Pratt}, {Pr{\'e}zeau}, {Prunet},
  {Puget}, {Rachen}, {Reach}, {Rebolo}, {Reinecke}, {Remazeilles}, {Renault},
  {Renzi}, {Ristorcelli}, {Rocha}, {Rosset}, {Rossetti}, {Roudier},
  {Rouill{\'e} d'Orfeuil}, {Rowan-Robinson}, {Rubi{\~n}o-Mart{\'\i}n},
  {Rusholme}, {Said}, {Salvatelli}, {Salvati}, {Sandri}, {Santos},
  {Savelainen}, {Savini}, {Scott}, {Seiffert}, {Serra}, {Shellard}, {Spencer},
  {Spinelli}, {Stolyarov}, {Stompor}, {Sudiwala}, {Sunyaev}, {Sutton},
  {Suur-Uski}, {Sygnet}, {Tauber}, {Terenzi}, {Toffolatti}, {Tomasi},
  {Tristram}, {Trombetti}, {Tucci}, {Tuovinen}, {T{\"u}rler}, {Umana},
  {Valenziano}, {Valiviita}, {Van Tent}, {Vielva}, {Villa}, {Wade}, {Wandelt},
  {Wehus}, {White}, {White}, {Wilkinson}, {Yvon}, {Zacchei}, \&
  {Zonca}}]{Planck15}
{Planck Collaboration}, {Ade}, P.~A.~R., {Aghanim}, N., {et~al.} 2016, \aap,
  594, A13

\bibitem[{{Platts} {et~al.}(2020){Platts}, {Prochaska}, \& {Law}}]{Platts20}
{Platts}, E., {Prochaska}, J.~X., \& {Law}, C.~J. 2020, \apjl, 895, L49

\bibitem[{{Platts} {et~al.}(2019){Platts}, {Weltman}, {Walters}, {Tendulkar},
  {Gordin}, \& {Kandhai}}]{Platts19}
{Platts}, E., {Weltman}, A., {Walters}, A., {et~al.} 2019, \physrep, 821, 1

\bibitem[{{Plotnikov} \& {Sironi}(2019)}]{Plotnikov19}
{Plotnikov}, I., \& {Sironi}, L. 2019, \mnras, 485, 3816

\bibitem[{{Popov} \& {Postnov}(2013)}]{Popov13}
{Popov}, S.~B., \& {Postnov}, K.~A. 2013, arXiv:1307.4924, arXiv:1307.4924

\bibitem[{{Prochaska} \& {Zheng}(2019)}]{Prochaska19}
{Prochaska}, J.~X., \& {Zheng}, Y. 2019, \mnras, 485, 648

\bibitem[{{Ravi} {et~al.}(2019){Ravi}, {Catha}, {D'Addario}, {Djorgovski},
  {Hallinan}, {Hobbs}, {Kocz}, {Kulkarni}, {Shi}, {Vedantham}, {Weinreb}, \&
  {Woody}}]{Ravi19}
{Ravi}, V., {Catha}, M., {D'Addario}, L., {et~al.} 2019, \nat, 572, 352

\bibitem[{{Rest} {et~al.}(2005){Rest}, {Stubbs}, {Becker}, {Miknaitis},
  {Miceli}, {Covarrubias}, {Hawley}, {Smith}, {Suntzeff}, {Olsen}, {Prieto},
  {Hiriart}, {Welch}, {Cook}, {Nikolaev}, {Huber}, {Prochtor}, {Clocchiatti},
  {Minniti}, {Garg}, {Challis}, {Keller}, \& {Schmidt}}]{Rest05}
{Rest}, A., {Stubbs}, C., {Becker}, A.~C., {et~al.} 2005, \apj, 634, 1103

\bibitem[{{Safarzadeh} {et~al.}(2020){Safarzadeh}, {Prochaska}, {Heintz}, \&
  {Fong}}]{Safarzadeh20}
{Safarzadeh}, M., {Prochaska}, J.~X., {Heintz}, K.~E., \& {Fong}, W.-f. 2020,
  arXiv, 2009.11735

\bibitem[{{Schechter} {et~al.}(1993){Schechter}, {Mateo}, \&
  {Saha}}]{Schechter93}
{Schechter}, P.~L., {Mateo}, M., \& {Saha}, A. 1993, \pasp, 105, 1342

\bibitem[{{Schlafly} \& {Finkbeiner}(2011)}]{Schlafly11}
{Schlafly}, E.~F., \& {Finkbeiner}, D.~P. 2011, \apj, 737, 103

\bibitem[{{Scholz} {et~al.}(2016){Scholz}, {Spitler}, {Hessels}, {Bogdanov},
  {Brazier}, {Camilo}, {Chatterjee}, {Cordes}, {Crawford}, {Deneva}, {Ferdman},
  {Freire}, {Kaspi}, {Lazarus}, {Lynch}, {Madsen}, {McLaughlin}, {Patel},
  {Ransom}, {Seymour}, {Stairs}, {Stappers}, {van Leeuwen}, \&
  {Zhu}}]{Scholz16}
{Scholz}, P., {Spitler}, L., {Hessels}, J., {et~al.} 2016, in AAS/High Energy
  Astrophysics Division \#15, AAS/High Energy Astrophysics Division, 105.03

\bibitem[{{Scholz} {et~al.}(2020){Scholz}, {Cook}, {Cruces}, {Hessels},
  {Kaspi}, {Majid}, {Naidu}, {Pearlman}, {Spitler}, {Bandura}, {Bhardwaj},
  {Cassanelli}, {Chawla}, {Gaensler}, {Good}, {Josephy}, {Karuppusamy},
  {Keimpema}, {Kirichenko}, {Kirsten}, {Kocz}, {Leung}, {Marcote}, {Masui},
  {Mena-Parra}, {Merryfield}, {Michilli}, {Naudet}, {Nimmo}, {Pleunis},
  {Prince}, {Rafiei-Ravandi}, {Rahman}, {Shin}, {Smith}, {Stairs}, {Tendulkar},
  \& {Vanderlinde}}]{Scholz20}
{Scholz}, P., {Cook}, A., {Cruces}, M., {et~al.} 2020, \apj, 901, 165

\bibitem[{{Shannon} \& {Ravi}(2017)}]{Shannon17}
{Shannon}, R.~M., \& {Ravi}, V. 2017, \apjl, 837, L22

\bibitem[{{Shannon} {et~al.}(2018){Shannon}, {Macquart}, {Bannister}, {Ekers},
  {James}, {Os{\l}owski}, {Qiu}, {Sammons}, {Hotan}, {Voronkov}, {Beresford},
  {Brothers}, {Brown}, {Bunton}, {Chippendale}, {Haskins}, {Leach},
  {Marquarding}, {McConnell}, {Pilawa}, {Sadler}, {Troup}, {Tuthill},
  {Whiting}, {Allison}, {Anderson}, {Bell}, {Collier}, {G{\"u}rkan}, {Heald},
  \& {Riseley}}]{Shannon18}
{Shannon}, R.~M., {Macquart}, J.~P., {Bannister}, K.~W., {et~al.} 2018, \nat,
  562, 386

\bibitem[{{Skrutskie} {et~al.}(2006){Skrutskie}, {Cutri}, {Stiening},
  {Weinberg}, {Schneider}, {Carpenter}, {Beichman}, {Capps}, {Chester},
  {Elias}, {Huchra}, {Liebert}, {Lonsdale}, {Monet}, {Price}, {Seitzer},
  {Jarrett}, {Kirkpatrick}, {Gizis}, {Howard}, {Evans}, {Fowler}, {Fullmer},
  {Hurt}, {Light}, {Kopan}, {Marsh}, {McCallon}, {Tam}, {Van Dyk}, \&
  {Wheelock}}]{Skrutskie06}
{Skrutskie}, M.~F., {Cutri}, R.~M., {Stiening}, R., {et~al.} 2006, \aj, 131,
  1163

\bibitem[{{Spitler} {et~al.}(2014){Spitler}, {Cordes}, {Hessels}, {Lorimer},
  {McLaughlin}, {Chatterjee}, {Crawford}, {Deneva}, {Kaspi}, {Wharton},
  {Allen}, {Bogdanov}, {Brazier}, {Camilo}, {Freire}, {Jenet},
  {Karako-Argaman}, {Knispel}, {Lazarus}, {Lee}, {van Leeuwen}, {Lynch},
  {Ransom}, {Scholz}, {Siemens}, {Stairs}, {Stovall}, {Swiggum},
  {Venkataraman}, {Zhu}, {Aulbert}, \& {Fehrmann}}]{Spitler14}
{Spitler}, L.~G., {Cordes}, J.~M., {Hessels}, J.~W.~T., {et~al.} 2014, \apj,
  790, 101

\bibitem[{{Spitler} {et~al.}(2016){Spitler}, {Scholz}, {Hessels}, {Bogdanov},
  {Brazier}, {Camilo}, {Chatterjee}, {Cordes}, {Crawford}, {Deneva}, {Ferdman},
  {Freire}, {Kaspi}, {Lazarus}, {Lynch}, {Madsen}, {McLaughlin}, {Patel},
  {Ransom}, {Seymour}, {Stairs}, {Stappers}, {van Leeuwen}, \&
  {Zhu}}]{Spitler16}
{Spitler}, L.~G., {Scholz}, P., {Hessels}, J.~W.~T., {et~al.} 2016, \nat, 531,
  202

\bibitem[{{Stritzinger} {et~al.}(2018){Stritzinger}, {Taddia}, {Burns},
  {Phillips}, {Bersten}, {Contreras}, {Folatelli}, {Holmbo}, {Hsiao},
  {Hoeflich}, {Leloudas}, {Morrell}, {Sollerman}, \&
  {Suntzeff}}]{Stritzinger17}
{Stritzinger}, M.~D., {Taddia}, F., {Burns}, C.~R., {et~al.} 2018, \aap, 609,
  A135

\bibitem[{{Tanenbaum} {et~al.}(1968){Tanenbaum}, {Zeissig}, \&
  {Drake}}]{Tanenbaum68}
{Tanenbaum}, B.~S., {Zeissig}, G.~A., \& {Drake}, F.~D. 1968, Science, 160, 760

\bibitem[{{Tavani} {et~al.}(2020){Tavani}, {Verrecchia}, {Casentini}, {Perri},
  {Ursi}, {Pacciani}, {Pittori}, {Bulgarelli}, {Piano}, {Pilia}, {Bernardi},
  {Addis}, {Antonelli}, {Argan}, {Baroncelli}, {Caraveo}, {Cattaneo}, {Chen},
  {Costa}, {Di Persio}, {Donnarumma}, {Evangelista}, {Feroci}, {Ferrari},
  {Fioretti}, {Lazzarotto}, {Longo}, {Morselli}, {Paoletti}, {Parmiggiani},
  {Trois}, {Vercellone}, {Naldi}, {Pupillo}, {Bianchi}, \&
  {Puccetti}}]{Tavani20}
{Tavani}, M., {Verrecchia}, F., {Casentini}, C., {et~al.} 2020, \apjl, 893, L42

\bibitem[{{Tendulkar} {et~al.}(2020){Tendulkar}, {Gil de Paz}, {Kirichenko},
  {Hessels}, {Bhardwaj}, {{\'A}vila}, {Bassa}, {Chawla}, {Fonseca}, {Kaspi},
  {Keimpema}, {Kirsten}, {Lazio}, {Marcote}, {Masui}, {Nimmo}, {Paragi},
  {Rahman}, {Reverte Pay{\'a}}, {Scholz}, \& {Stairs}}]{Tendulkar20}
{Tendulkar}, S.~P., {Gil de Paz}, A., {Kirichenko}, A.~Y., {et~al.} 2020, arXiv
  e-prints, arXiv:2011.03257

\bibitem[{{Thornton} {et~al.}(2013){Thornton}, {Stappers}, {Bailes},
  {Barsdell}, {Bates}, {Bhat}, {Burgay}, {Burke-Spolaor}, {Champion}, {Coster},
  {D'Amico}, {Jameson}, {Johnston}, {Keith}, {Kramer}, {Levin}, {Milia}, {Ng},
  {Possenti}, \& {van Straten}}]{Thornton13}
{Thornton}, D., {Stappers}, B., {Bailes}, M., {et~al.} 2013, Science, 341, 53

\bibitem[{{Wadiasingh} \& {Timokhin}(2019)}]{Wadiasingh19}
{Wadiasingh}, Z., \& {Timokhin}, A. 2019, \apj, 879, 4

\bibitem[{{Yamasaki} {et~al.}(2016){Yamasaki}, {Totani}, \&
  {Kawanaka}}]{Yamasaki16}
{Yamasaki}, S., {Totani}, T., \& {Kawanaka}, N. 2016, \mnras, 460, 2875

\bibitem[{{Zampieri} {et~al.}(2020){Zampieri}, {Burtovoi}, {Fiori}, {Naletto},
  {Ochner}, {Turatto}, {Nicastro}, {Palazzi}, {Pilia}, {Possenti}, {Casella},
  \& {Rea}}]{Zampieri20}
{Zampieri}, L., {Burtovoi}, A., {Fiori}, M., {et~al.} 2020, The Astronomer's
  Telegram, 13493, 1

\end{thebibliography}

\end{document}